\documentclass[12pt, draftclsnofoot, onecolumn]{IEEEtran}

\usepackage{graphicx} 
\graphicspath{{Fig/}}
 
\usepackage{graphicx}
\usepackage{amsmath,bbm,amssymb,amsfonts,amstext,amsopn,sansmath}
\usepackage{cite}
\usepackage{balance}
\usepackage{url}
\usepackage{epsfig}
\usepackage{setspace}
\usepackage{stmaryrd}
\usepackage{psfrag}	
\usepackage{multirow}
\usepackage{makecell}
\usepackage{float}
\usepackage[process=auto]{pstool}
\usepackage{etoolbox}
\usepackage{algorithm}
\usepackage{algorithmic}
\allowdisplaybreaks
\usepackage{textcomp}
\usepackage{multicol}
\usepackage{xfrac}
\usepackage{colortbl}

\usepackage{clipboard}
\newclipboard{Paper_V1}
 
\usepackage{xcolor}
\makeatletter
\let\myorg@bibitem\bibitem
\def\bibitem#1#2\par{%
	\@ifundefined{bibitem@#1}{%
		\myorg@bibitem{#1}#2\par
	}{%
		\begingroup
		\color{\csname bibitem@#1\endcsname}%
		\myorg@bibitem{#1}#2\par
		\endgroup
	}%
}
\newcommand{\highlightref}[1]{\expandafter\newcommand\expandafter*\csname bibitem@#1\endcsname{blue}}
\makeatother

\highlightref{zheng2019intelligent}

\newcommand{\Trans}{\mathsf{T}}
\newcommand{\e}{\mathsf{e}}

\def\bs{\mathbf{s}}
\def\by{\mathbf{y}}
\def\bA{\mathbf{A}}

\def\bx{\mathbf{x}}
\def\bn{\mathbf{n}}
\def\bV{\mathbf{V}}

\def\bq{\mathbf{q}}
\def\bone{\boldsymbol{1}}
\def\bzero{\boldsymbol{0}}

\def\bp{\mathbf{p}}

\def\bA{\mathbf{A}}

\def\bu{\mathbf{u}}

\def\ba{\mathbf{a}}
\def\bM{\mathbf{M}}
\def\bw{\mathbf{w}}
\def\bX{\mathbf{X}}

\def\byrv{\boldsymbol{\mathnormal{y}}}
\def\bxrv{\boldsymbol{\mathnormal{x}}}

\def\sR{\mathbb{R}}
\def\sRp{\sR_{\geq0}}

\def\sQ{\mathcal{Q}}
\def\sM{\mathcal{M}}
\def\sX{\mathcal{X}}

\def\Ex{\mathbb{E}} 
\def\Vx{\mathbb{V}}

\def\cA{\mathcal{A}}
\def\cAc{\mathcal{A}^{\rm c}}

\def\cM{\mathcal{M}}

\newcommand{\diag}{\mathrm{diag}}

\newcommand{\thr}{\mathrm{thr}}

\newcommand{\argmax}{\mathrm{argmax}}

\newcommand{\rls}{\mathrm{rls}}

\newtheorem{remk}{Remark}
\newtheorem{defin}{Definition}

\newtoggle{OneColumn}
\toggletrue{OneColumn}

\newtoggle{arXiv}
\togglefalse{arXiv}

\usepackage{lipsum}
\def\@IEEEBIOphotowidth{1cm}    
\def\@IEEEBIOphotodepth{1cm}   
\def\@IEEEBIOhangwidth{1.2cm}    
\def\@IEEEBIOhangdepth{1.2cm}    

\EndPreamble
\begin{document}
\title{Olfaction-inspired MCs: Molecule Mixture Shift Keying and Cross-Reactive Receptor Arrays \vspace{-0.01cm}}


\author{Vahid Jamali, Helene M. Loos, Andrea Buettner,  Robert Schober, and \\  H. Vincent Poor  \vspace{-0.1cm}
	\thanks{Vahid Jamali's work was funded by the Deutsche Forschungsgemeinschaft (DFG -- German Research Foundation) under project number JA 3104/1-1.}
\thanks{V. Jamali and H. V. Poor  are with the Department of Electrical and Computer Engineering, Princeton University, Princeton, NJ 08544 USA (e-mail: \{jamali, poor\}@princeton.edu).}
\thanks{H. M. Loos and A. Buettner are with the Chair of Aroma and Smell Research at Friedrich-Alexander University Erlangen-N\"urnberg (FAU), Erlangen 91054, Germany (e-mail: \{helene.loos,andrea.buettner\}@fau.de) and the Fraunhofer Institute for Process Engineering and Packaging,  Freising 85354, Germany (e-mail: \{helene.loos,andrea.buettner\}@ivv.fraunhofer.de).}
\thanks{R. Schober is with the Institute for Digital Communications at Friedrich-Alexander University Erlangen-N\"urnberg (FAU), Erlangen 91058, Germany (e-mail: robert.schober@fau.de).}
}

\maketitle

\vspace{-1cm}

\begin{abstract}
In this paper, we propose a novel concept for engineered molecular communication (MC) systems inspired by animal olfaction. We focus on a multi-user scenario where several  transmitters wish to communicate with a central receiver. We assume that each transmitter employs a unique mixture of different types of signaling molecules to represent its message and  the receiver is equipped with an array comprising $R$ different types of receptors in order to detect the emitted molecule mixtures.  The design of an MC system based on \textit{orthogonal} molecule-receptor pairs implies that the hardware complexity of the receiver  linearly scales  with the number of signaling molecule types $Q$ (i.e., $R=Q$). Natural olfaction systems avoid such high complexity by employing  arrays of \textit{cross-reactive} receptors, where each type of molecule activates multiple types of receptors and each type of receptor is predominantly activated by multiple types of molecules albeit with different activation strengths. For instance, the human olfactory system is believed to discriminate several thousands of chemicals using only a few hundred receptor types, i.e., $Q\gg R$. Motivated by this observation, we first develop an end-to-end MC channel model that accounts for the key properties of olfaction. Subsequently, we present the proposed transmitter and receiver designs. In particular, given a set of signaling molecules, we develop algorithms that allocate molecules to different  transmitters and optimize the mixture alphabet  for communication. Moreover, we formulate the molecule mixture recovery as a convex compressive sensing problem which can be efficiently solved via available numerical solvers.  Finally, we present a comprehensive set of simulation results to evaluate the performance of the proposed MC designs revealing interesting  insights regarding the design parameters. For instance, we show that mixtures comprising few types of molecules are best suited for communication since they can be more reliably detected by the cross-reactive array than one type of molecule or mixtures of many molecule types.\vspace{-0.1cm}
\end{abstract}

\begin{IEEEkeywords}
Engineered molecular communication, transmitter and receiver design, molecule mixture modulation, compressive sensing, odor, pheromone, and generalist receptors.
\end{IEEEkeywords}
\vspace{-0.3cm}

\section{Introduction}

Unlike the widely-used electromagnetic-based communication (EMC) systems which embed data into the properties of electromagnetic waves (e.g., amplitude, phase, and frequency), molecular communication (MC) systems encode information into the features of  molecular signals (e.g., concentration, identity, and time  and dynamics of release) \cite{nakano2013molecular,farsad2016comprehensive,jamali2019channel}. Examples of MC in nature are abundant ranging from cell-to-cell communications \cite{alberts2015essential,hartmann2013sperm,veitinger2017ectopic}, to hormone communications via the bloodstream \cite{norman2015hormones}, and long-range pheromone communications among animals and plants \cite{wyatt2003pheromones,kornbausch2022odorant}. Synthetic MC has recently emerged as a new concept in communication engineering with prospective revolutionary applications in medicine, environmental engineering, and manufacturing. For example, in medical applications, MC systems are expected to facilitate distributed nano-/micro-scale sensor networks for detection of cancer cells and  the control of actuators for targeted drug release \cite{chude2017molecular,soldner2020survey}. Furthermore, MC can be deployed in industrial settings where traditional EM-based communication may be unsafe (e.g., in chemical reactors processing explosive gases) or inefficient (e.g., in oil pipes) \cite{farsad2016comprehensive}.

\subsection{State-of-the-Art MC Designs and Associated Challenges}

Due to their inherent differences, EMC and MC systems present different challenges and require different design considerations. For example, MC is much slower than EMC due to the slower propagation of mass compared to waves. More importantly, the released molecules
stay in the channel for a long time and constitute interference for the detection of molecules released  by the same transmitter (Tx) or other Txs at future times. These phenomena fundamentally limit how frequently and how reliably the MC channel can be used. The problem of interference mitigation has been extensively studied in the MC literature and various countermeasures have been proposed including equalization at the receiver (Rx) \cite{tepekule2015isi,kilinc2013receiver},  deployment of enzymes in the channel \cite{noel2014improving}, and  use of multiple reactive signaling molecules (e.g., acids and bases) by the Tx \cite{farsad2017novel,jamali2018diffusive}. However, these solutions have limited applicability and efficiency. For example,  equalization techniques are typically quite complex and sensitive to channel estimation and synchronization errors. Similarly, the use of enzymes is limited to specific application scenarios such as the small space between nerve and muscle cells and the use of acids/bases is limited
to specific water-based solutions and cannot be easily applied in other fluid or air based environments.  

Interference and slow molecule propagation impact the performance of  concentration-shift keying (CSK) and time of release-shift keying (TSK) modulations more severely than that of  molecule-shift keying (MSK) modulation \cite{kuscu2019transmitter}. In fact,  in principle, if the Rx is able to detect a large number of different types of signaling molecules, then using MSK modulation, Txs may access the MC channel frequently by releasing  different types of molecules into the channel while sufficiently separating consecutive releases of the same type of molecule  in order to reduce interference. However, the MSK-based MC designs proposed so far in the literature \cite{jamali2018diffusive,kuscu2019transmitter,kuran2011modulation,shahmohammadian2012optimum,chen2020generalized} employ only a small number of different types of molecules, denoted by $Q$,  and are not scalable for large $Q$ due to the following limitations: 
\begin{itemize}
	\item First, it is implicitly assumed in MSK modulation that for each adopted signaling molecule type, the Rx is equipped with a corresponding specifically-tuned receptor type\footnote{Throughout this paper, we use the term \textit{receptor} to refer to the sensing unit at the Rx which maybe a biological receptive structure (e.g., a receptor or a binding protein) or a non-biological sensor.}. This implies that the Rx architecture complexity linearly increases with $Q$ which limits the feasibility of this Rx architecture for large $Q$. 
	\item Second, the ideal assumption that each receptor type \textit{only} responds to the corresponding signaling molecule type  simplifies the design of the MC system but  may not be valid in practice \cite{tisch2010nanomaterials}. Most receptors respond to different chemical substances  albeit with different strengths. This ideal assumption becomes particularly limiting when a large set of molecule and receptor types are considered.
	\item Third, the detection algorithms designed for small $Q$ may not be directly applicable for large $Q$ as their complexity may not scale favorably with $Q$.  
\end{itemize} 

For future reference, we refer to  MC systems that employ a Dedicated
Receptor type for EAch Molecule type  as DREAM systems.

\begin{figure}[t]
	\centering
	\includegraphics[width=0.6\columnwidth]{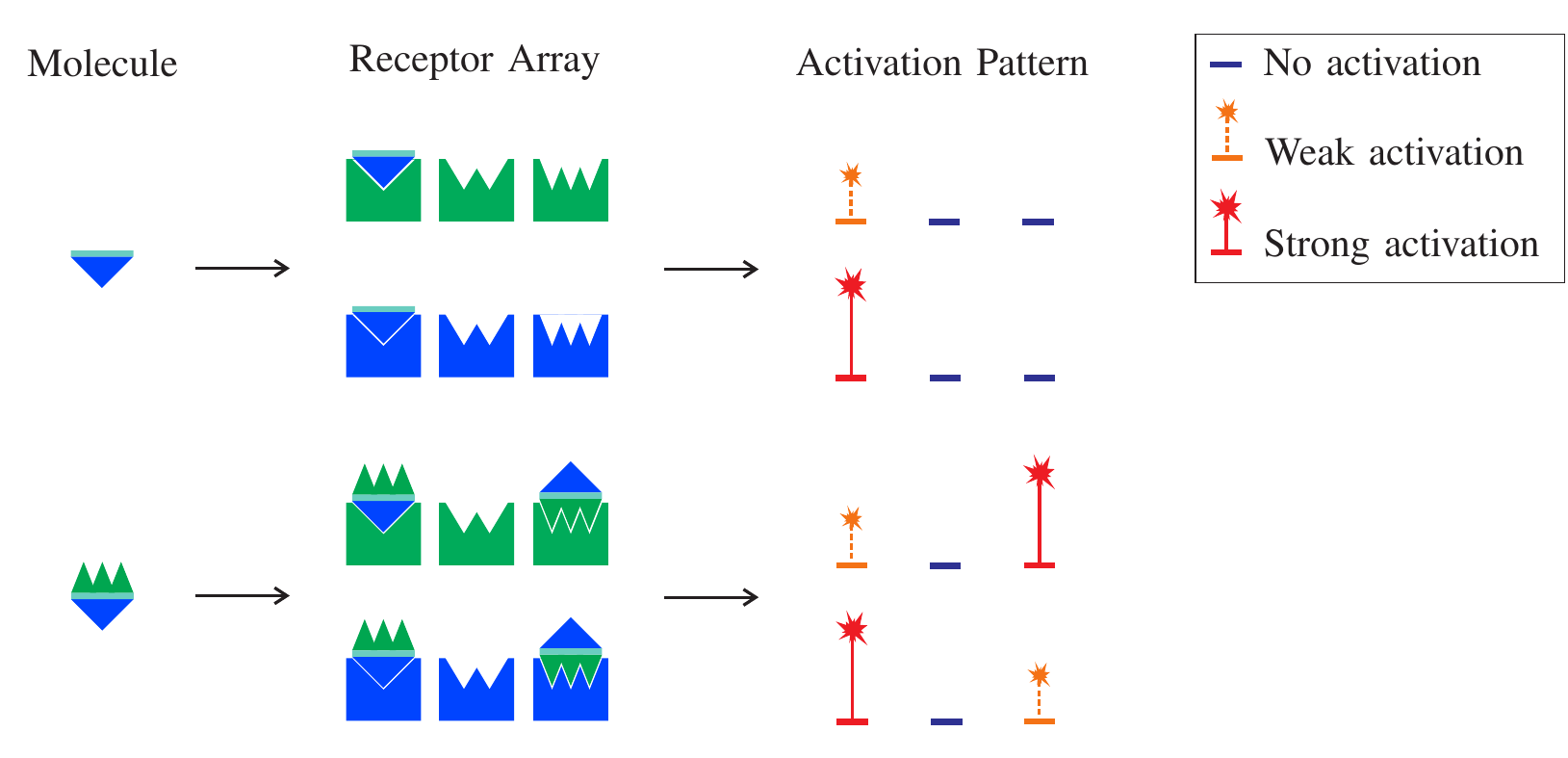}\vspace{-0.3cm}
	\caption{Schematic illustration of the cross-reactive receptor arrays of  olfactory systems known as \textit{shape-pattern theory of olfactory perception} \cite{buck2005unraveling,Pillow2017}. In this illustration, each molecule type activates those receptor types that have a matching geometrical shape and the strength of activation is enhanced if they have identical color. } \label{Fig:Reception} \vspace{-0.3cm}
\end{figure} 

\subsection{Proposed Olfaction-inspired MC}

Fortunately, nature offers several solutions for the concurrent detection of a large number of  different types of molecules, which can be found in the chemosensory systems of different organisms. In fact, for airborne chemicals, the olfactory systems of mammals and insects can detect molecular information in an efficient manner and have vital functions in, e.g., navigating, foraging, and reproduction  \cite{buck2005unraveling,su2009olfactory,kaupp2010olfactory}. For instance, it
is estimated that humans are able to perceive $\sim 10^4-10^5$ different chemicals as having  distinct odors\footnote{The olfactory system is able to recognize various \textit{odorants}. Odorants belong to large chemical substance classes with a high diversity of chemical features that share a certain degree of volatility and both polar and nonpolar properties. They can be as diverse as being esters, alcohols, thiols,  terpenoids, and aromatic substances to name a few  \cite{buettner2017springer,su2009olfactory}. The animal olfactory system is, among other systems, also involved in intra-species communication via \textit{pheromones}. Pheromones are chemical molecule substances that are used for communication among the individuals of a species and can regulate their behavior \cite{wyatt2003pheromones}. Furthermore, the general term \textit{volatile organic compounds (VOCs)} (volatiles  are chemical substances easily evaporated at room temperature) has been widely used in the literature  to refer to the chemical cue molecules detected by  animal olfactory systems  \cite{pearce2006handbook,buettner2017springer}. In this paper, we adopt a broad view of chemicals representing information and refer to them as signaling molecules.} using only $\sim 400$ different types of olfactory receptors (ORs)\footnote{The Nobel Prize in Physiology and Medicine in 2004 was awarded jointly to Richard Axel and Linda B. Buck for their discoveries of ``odorant receptors and the organization of the olfactory system`` \cite{buck2005unraveling,buck1991novel}.}  \cite{buck2005unraveling}.  Unlike DREAM systems where each receptor type is assumed to be narrowly tuned to its respective type of signaling molecule, in these natural systems, the majority of the receptors are broadly tuned\footnote{In addition to  widely-tuned receptor types (known as generalist), natural olfactory systems are equipped with so-called specialist receptor types that are narrowly tuned to specific chemical substances of particular importance (e.g.,  predator or mate odor) \cite{dunkel2014nature}. In this paper, we mainly focus on generalist receptor types which enable the discrimination of a large number of signaling molecule types.} to multiple types of signaling molecules while most signaling molecule types have affinity with multiple types
of receptors. Therefore, the odor discriminatory capacity of the olfactory system stems from an array of \textit{cross-reactive} receptor types, which extracts the information regarding the presence of  molecules and encodes it into the \textit{activation pattern} of the ORs, see Fig.~\ref{Fig:Reception}. Over the past decades, natural olfaction not only has received extensive attention in biology, neuroscience, and other natural sciences but has also served as a bio-inspired model for synthetic sensor design (known as artificial olfaction or bioelectronic nose) in biotechnology, material science, etc. \cite{barbosa2018protein,dung2018applications,pelosi2018gas}.  However, to the best of the authors' knowledge, the design and analysis of an engineered MC system inspired by natural olfaction, and more specifically, based on a cross-reactive receptor array, has not been considered in the MC literature, yet. We note that the multiple-input multiple-output (MIMO) MC systems studied in the literature \cite{meng2012mimo,koo2016molecular,gursoy2019index} employ multiple receptors of the \textit{same} type and a \textit{single} type of molecule and their generalization to MSK modulation suffers from the same limitations as the DREAM MC systems discussed in the previous subsection.

The main objective of this paper is to investigate the potential benefits of exploiting the properties of natural olfaction for the design of engineered MC systems. As a concrete case study, we focus on a sensor network (e.g., in an industrial setting) where each sensor (as a Tx) has occasionally a small amount of data to report to a central unit (as the Rx). Synthetic MC systems may be a preferred choice as they avoid the potential safety concerns of  traditional EMC in these applications and offer more secure communication since the molecular signal cannot be intercepted as easily as  EM waves.  Moreover, in such applications, the overhead needed for establishing 
synchronous transmission among multiple Txs can be large, and hence,  random multiple access is preferred. We assume that each Tx employs a unique mixture of different types of signaling molecules to represent its message and the Rx is equipped with an array of $R$ types of cross-reactive receptors in order to identify the emitted molecule mixtures. We refer to the proposed modulation scheme as \textit{molecule mixture shift keying (MMSK)}. This paper makes the following contributions:  

\begin{itemize}
	\item We first develop a communication-theoretical model that captures the key properties of olfaction including asynchronous molecule mixture signaling and cross-reactive receptor arrays. To this end, we present a brief review of the natural olfaction process and the state-of-the-art in artificial olfaction. Subsequently, we identify a set of key concepts and incorporate them in the proposed communication-theoretical model.  
	\item Given a set of $Q$ types of molecules, we first develop an algorithm that allocates these molecules to Txs such that each molecule is assigned to one Tx. Subsequently, we develop another algorithm that constructs the best mixture alphabet for communication for each Tx. To do so, we define a \textit{dissimilarity} metric that quantifies how dissimilar/distinguishable two  mixtures of different types of molecules are based on the Rx array output. Using this metric, the first algorithm allocates those molecule types to a given Tx  that are most dissimilar  since these molecules are released simultaneously in mixtures by that Tx. Then, for each Tx, the second algorithm constructs a set of mixtures (as the alphabet of the proposed MMSK modulation) that are most distinguishable by the Rx.	
	\item While the Rx has to discriminate the large number of molecule types used by all Txs for signaling, the number of molecule types that are concurrently present around the Rx is typically small due to the assumption of  sporadic channel access by the Txs\footnote{A similar analogy exists for natural olfactory systems, where the total number of perceivable odors is large, but the number of simultaneously perceivable odors is small \cite{thomas2014perception}.}. 
	Therefore, we formulate the molecule mixture recovery at the Rx as a compressive sensing problem which explicitly exploits the knowledge of the MMSK modulation alphabets used by Txs. Moreover, due to its convexity, the proposed recovery problem can be efficiently solved via existing numerical solvers. The recovery problem does not assume knowledge of the identity of the active Tx; however, this knowledge can be inferred due to the proposed unique molecule-Tx association. Therefore, we extend the proposed recovery problem to selectively process the received signal after first inferring the identity of the active Tx. Furthermore, we propose a matched filter that combines observations for consecutive sampling times to enhance the overall recovery performance. 	
	\item Finally, we present a comprehensive set of simulation results in order to evaluate the performance of the proposed MC designs.  These results reveal interesting  insights regarding the choices of the design parameters. For example, we show that in principle, it is not optimal to use only one type of molecule or mixtures of many types of molecules for signaling, and the optimal number of constituent molecule types used to construct signaling mixtures is numerically determined for an example scenario.  
\end{itemize}

The remainder of this paper is organized as follows. In Section~II, some preliminaries on natural and artificial olfaction as well as the proposed end-to-end channel model are presented.  The proposed transmitter and receiver designs are provided in Sections~III-A and III-B, respectively. Simulation results are reported in Section~IV, and conclusions are drawn in Section~V.

\textit{Notations:} Bold small letters (e.g., $\bx$) and bold capital letters  (e.g., $\bX$) denote vectors and matrices, respectively. Sets are shown by calligraphic letters (e.g., $\sX$). $\bX^{\Trans}$ represents the transpose of matrix $\bX$ and $[\bX]_{n,m}$ denotes the entry on the $n$-th row and $m$-th column of matrix $\bX$ where sub-index $m$ is dropped for vectors. We use $\by = |\bx|$ to denote  element-wise absolute value, i.e., $[\by]_{n}=|[\bx]_{n}|$. $\|\bx\|_p$ denotes the vector $p$-norm and $\bone_{n}$ and $\bzero_n$ are the all-one and all-zero vectors of size $n$, respectively. $|\sX|$ denotes the cardinality of set $\sX$.  $\sR$ and $\sRp$ are the sets of real and non-negative real numbers, respectively. ${\rm Pois}(\lambda)$ represents a Poisson random variable (RV) with mean $\lambda$.  $\Ex\{\bX\}$ and $\Vx\{\bX\}$ denote the statistical expectation and the variance of the entries of matrix $\bX$, respectively.

\begin{figure}[t]
	\centering
	\includegraphics[width=0.55\columnwidth]{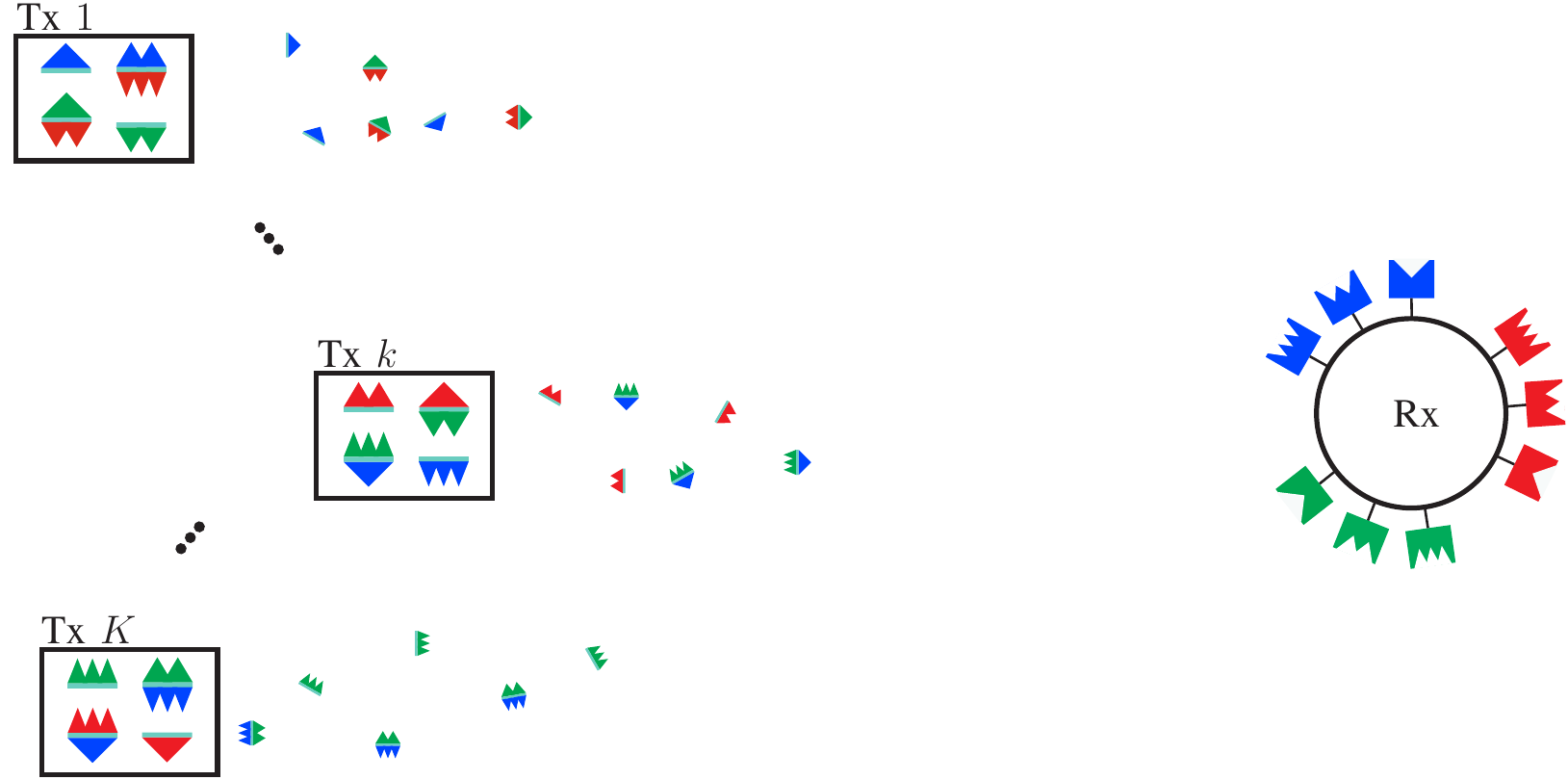}\vspace{-0.1cm}
	\caption{Proposed multi-user MC system consisting of several Txs that employ MMSK modulation to convey their messages to an Rx equipped with a cross-reactive receptor array.   \vspace{-0.3cm}} \label{Fig:MultiuserSystem} \vspace{-0.3cm}
\end{figure} 

\section{Communication-theoretical End-to-End Channel Model}\label{sec:channel}

We consider a multi-user MC system consisting of $K$ Txs  sending  messages to an Rx, e.g., as a model of a sensor network, see Fig.~\ref{Fig:MultiuserSystem}. We assume that the Txs employ MMSK modulation, namely each Tx releases a mixture of different molecule types to represent its message, cf. Section~\ref{sec:Tx}. We assume each Tx sporadically accesses the MC channel in a random access manner since it has only occasionally a small amount of data to transmit (e.g., reporting a change in temperature measured by a  sensor) which does not justify the establishment of synchronous communication. The released molecules propagate through the MC channel (e.g., via diffusion, turbulent flow, etc.) and some reach the Rx. The Rx is equipped with an array of $R$ cross-reactive types of receptors and recovers the transmitted messages and the identity of the corresponding Txs by processing the array signals. The objective of this section is to develop an end-to-end channel model that relates the  mixture signals released by the Txs to the  received array signals at the Rx. To do so, in the following, we first review the basic properties of olfaction, which form the basis for the proposed communication-theoretical channel model.

\subsection{Basic Properties of Olfaction}\label{sec:olfaction}
We first establish some key properties of natural olfaction. Then, we discuss how these properties have been exploited in sensor technology to develop olfaction-inspired gas sensors, which may serve as a potential building block for the Rx of the proposed synthetic MC system.

%

\subsubsection{\textbf{Natural olfaction}} 
\begin{figure}[t]
	
	\begin{minipage}{0.58\textwidth}\vspace{-0.9cm}
		\centering
		\includegraphics[width=0.98\columnwidth]{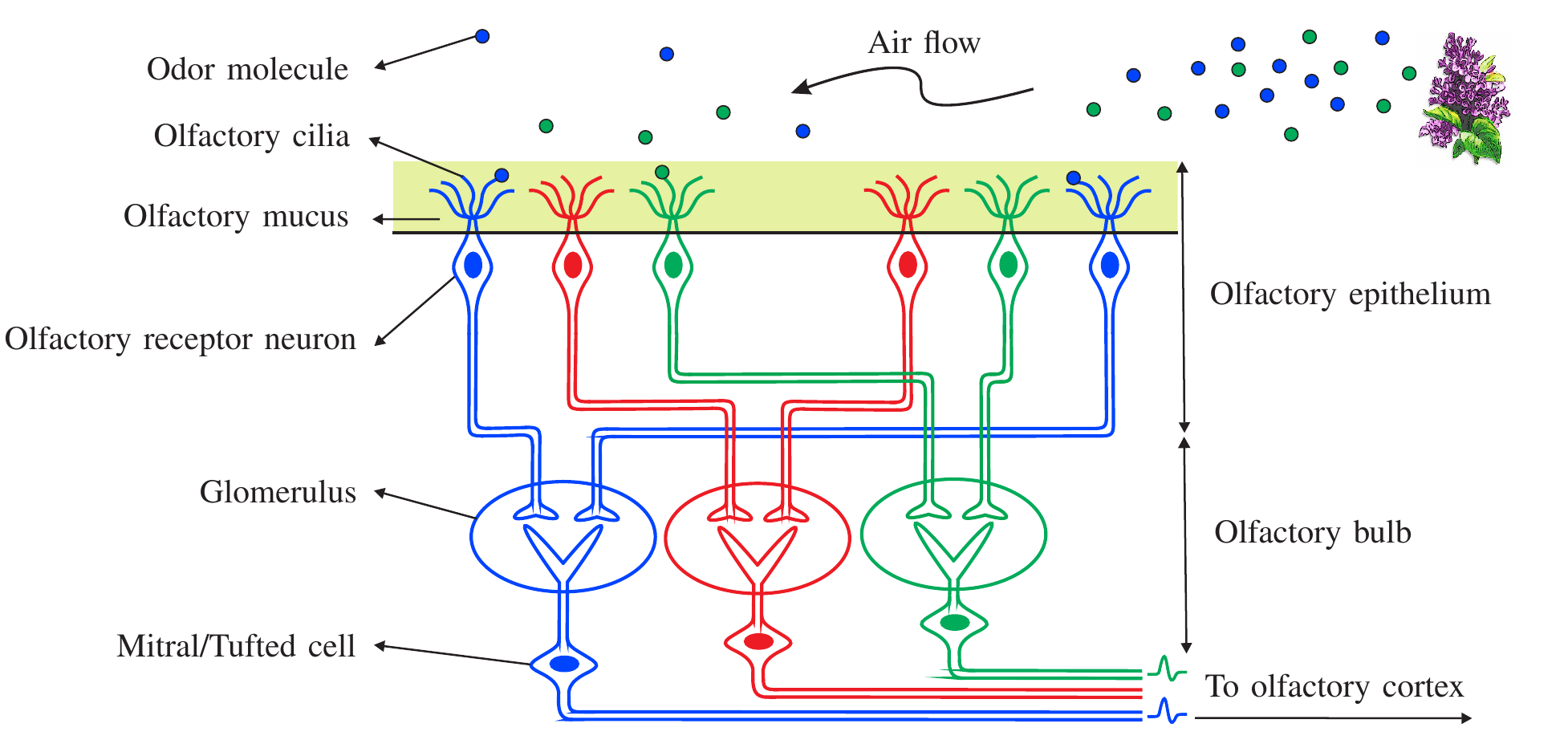}\vspace{-0.1cm}
		\caption{Schematic illustration of the signal flow in the mammalian olfactory system consisting of olfactory epithelium, olfactory bulb, and olfactory~cortex.   \vspace{-0.3cm}} \label{Fig:Olfactory} \vspace{-0.3cm}
	\end{minipage}
	\begin{minipage}{0.01\textwidth}
		\quad
	\end{minipage}
	\begin{minipage}{0.41\textwidth} 
		\centering
		\includegraphics[width=0.83\columnwidth]{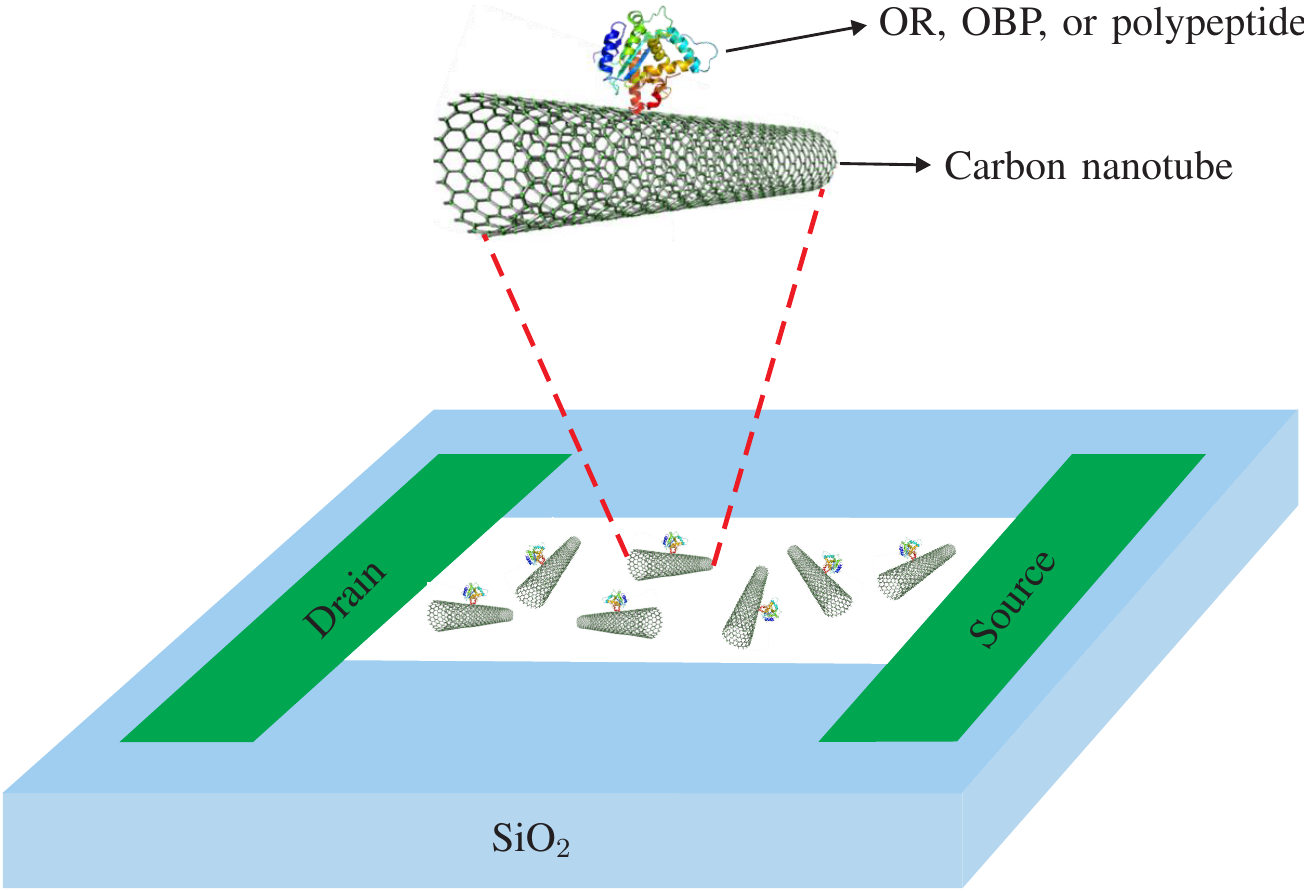}\vspace{0.02cm}
		\caption{Schematic illustration of a biosensor employing a carbon nanotube field-effect transistor functionalized with ORs, OBPs, or polypeptides (inspired by \cite{son2016bioelectronic}).} \label{Fig:Enose} \vspace{-0.3cm}
	\end{minipage}
	
\end{figure}

The perception of odor molecules, e.g., in humans, is initiated by several \textit{millions} of olfactory receptor neurons  located on the wall of the nasal cavity in the olfactory epithelium \cite{mori1995molecular}. In particular, each odorant receptor (OR) neuron has one dendrite from which $\sim15$ cilia extend into the nasal mucus. Odor molecules that reached the nasal cavity are dissolved in the nasal mucus and interact with the receptor proteins on the surface membrane of the cilia\footnote{The odor molecules either  interact directly with ORs or indirectly via odor-binding proteins (OBPs). In the latter case, which is more common in insect olfactory systems, the molecules bind to OBPs (in the nasal mucus of mammals or in the sensilla lymph of insects) which carry them to the ORs \cite{kaupp2010olfactory,leal2013odorant}.}. The ORs belong to a large  multigene family of G protein-coupled receptors (GPCRs), e.g., $\sim400$ types of ORs in humans and $\sim1000$ in mice. It is believed that each OR neuron expresses mainly one OR gene \cite{buck1991novel}. OR proteins vary extensively in their amino acid sequences which is how they interact with odorants with different structures \cite{buck2005unraveling}. In a nutshell, the animal olfactory system can be seen as a large array of cross-reactive receptors.  The binding of odor molecules to OR proteins triggers a cascade of signal transductions that
eventually open the cation channels in the OR neuron membrane which depolarizes the OR neuron and generates an action potential \cite{mori1995molecular}. Each OR neuron has a single axon that extends to the olfactory bulb. The OR neuron axons enter spherical structures, called glomeruli, and synapse with the dendrites of bulb neurons. It is known that each glomerulus receives signals from a \textit{single type} of OR and that each type of OR sends signals to only few ($\sim2-4$) glomeruli \cite{mori1995molecular,laurent1999systems}. In other words, the identities of the OR types are preserved at the olfactory bulb. Each mitral/tufted cell in the olfactory bulb receives input from a single glomerulus and relays it to the olfactory cortex for processing and eventually perception of the odor. Fig.~\ref{Fig:Olfactory} schematically illustrates the signal flow in the mammalian olfactory~system.

\subsubsection{\textbf{Artificial olfaction}} The extraordinary discriminatory power of natural olfactory systems to perceive a vast number of volatile organic compounds (VOCs) even at very low concentrations has motivated researchers to develop biosensors that mimic the structural and functional features of natural olfaction \cite{pearce2006handbook}. These sensors consist of two main components, namely an array of cross-reactive sensing elements for chemical detection and a transduction unit that generates a processable (electrical, optical, ...) signal. Sensing elements may consist of non-biological materials (e.g., metal oxides, conductive polymers, and graphene \cite{pelosi2018gas}), biological components (e.g., ORs, OBPs, and polypeptides \cite{barbosa2018protein}), and hybrid components (e.g., immobilized ORs on a carbon graphene sheet   \cite{dung2018applications}). The interaction of sensing elements with VOCs results in the variation of their chemical and physical properties and is measured by the transduction unit. Transduction may be electrical (e.g., change in the resistance of sensing elements), mechanical (e.g., change in the resonance frequency of sensing elements), or optical (e.g., change in the refractive index of sensing elements) \cite{barbosa2018protein}.  As an example, Fig.~\ref{Fig:Enose} illustrates a biosensor, where biological sensing elements are immobilized on a carbon nanotube (CNT) and are placed between  source and drain of a field-effect transistor (FET). Upon binding of an odor molecule, the resistance of the CNT changes which can be measured by the current-voltage relation of  drain and source of the FET. Although olfactory-inspired biosensors are still an active area of research, various small- and large-scale sensor arrays have been already reported in the literature \cite{lorwongtragool2014novel,janzen2006colorimetric,bernabei2012large}.

\subsection{Communication-theoretical Channel  Model}

In the following, we develop an abstract communication-theoretical channel model for the proposed olfactory-inspired MC systems. We note that our objective is not to develop an accurate model for natural olfaction or any particular artificial olfactory system. Instead, we propose a generic model that captures some \textit{key properties} of these systems. 

\subsubsection{\textbf{Key properties of olfaction}}\label{sec:olfprop} Based on the descriptions given in Section~\ref{sec:olfaction}, we identify the following key properties of olfactory systems: 
\begin{itemize}
	\item \textbf{P1 (cross-reactive receptor array):} The main reason for the high discriminatory capacity of olfactory systems is their cross-reactive receptor array structure. In other  words, one type of molecule  may activate multiple types of receptors and conversely, one type of receptor can be activated by different types of molecules \cite{buck2005unraveling,kaupp2010olfactory,mori1995molecular}.
	\item \textbf{P2 (sparsity):} Although the olfactory system is
	able to distinguish thousands of different molecule types, the number of molecule types that it can simultaneously identify is limited. For instance, various experiments demonstrated that humans can hardly distinguish $3-4$ odorants in a mixture that contains up to $8$ odorants \cite{thomas2014perception}. Mathematically speaking, successful recovery mandates the sparsity of the received signal. 
	\item \textbf{P3 (concentration):} Higher concentrations of molecules activate more types of receptors because the activation of some types of receptors  is concentration dependent requiring a minimum concentration of molecules to proceed with detection\footnote{The natural olfactory system has the capability of adapting its detection sensitivity to the  intensity of  stimulation  (known as odor adaptation) which is essential for preventing saturation and allowing the retention of high detection sensitivity during continuous odor stimulation \cite{zufall2000cellular}.}. This suggests that the concentrations  and identities of the molecules are encoded in the activation pattern of different types of receptors\footnote{In addition to the spatial activation pattern of the ORs, the odor concentration is believed to be encoded into the temporal response of OR neurons \cite{bolding2017complementary}. However, for simplicity, we neglect the concentration information contained in the temporal response in this paper.} \cite{su2009olfactory,bolding2017complementary}.
	\item \textbf{P4 (inhibition):} A given type of molecules can activate some types of receptors
	and inhibit others, whereas an individual receptor can be activated by some types of molecules and inhibited by others \cite{su2009olfactory}. While various examples of receptor inhibition have been discussed in the literature, see, e.g., \cite{hallem2004molecular,pfister2020odorant},  inhibition  is less common (or at least not sufficiently understood) compared to  excitation/activation  particularly in vertebrates\footnote{In addition to the odor-invoked inhibition of OR neurons \cite{pfister2020odorant}, the glomeruli and mitral/tufted cells are locally inhibited via so-called periglomerular (PG) cells and granule (Gr) cells, respectively, which is known as lateral inhibition \cite{mori1995molecular,laurent1999systems,kaupp2010olfactory}. These cells are believed to perform a sort of local processing in the olfactory bulb which enables finer odor contrast/discrimination by adaptively narrowing the receptive range. This form of inhibition is not modeled in this paper.} \cite{kaupp2010olfactory}.  
	\item \textbf{P5 (noise):} OR neurons are noisy and spontaneously active (i.e., they fire action potentials even in the absence of an odorant). For instance, the OR neuron of a Drosophila fly fires $\sim8$ spikes/s in the absence of odor \cite{wilson2013early}. The aggregation of the signal from all OR neurons of the same type yields thousands of noisy baseline spikes/s. 
	\item \textbf{P6 (sensitivity enhancement):} Each glomerulus receives signals from thousands of OR neurons of the same type. This signal aggregation enhances the detection sensitivity and the signal-to-noise ratio (SNR) of the signal sent to the olfactory cortex  by averaging out of the uncorrelated noises across the distributed receptors~\cite{laurent1999systems}.	
\end{itemize}

In our communication model, we mainly focus on modeling the Rx reception properties \textbf{P1}-\textbf{P6}. As is elaborated in the following, for the Tx model, we assume \textit{random} and \textit{asynchronous} release, and for the MC channel, we adopt a generic model capturing the dispersive nature of molecule propagation.  

\subsubsection{\textbf{End-to-end channel model}} The end-to-end channel model accounts for the release, propagation, and reception mechanisms of the system. In particular, as  channel inputs, we consider  mixtures of different types of signaling molecules released into the channel by the Txs, where each molecule mixture coveys a certain message (e.g., a change in the temperature measured by a sensor). Let $\sQ$ and $\sM$ denote the sets of all molecule types and molecule mixtures used by the Txs, respectively, where $Q\triangleq|\sQ|$ and $M\triangleq|\sM|$. The transmit signal vector is defined as $\bs=[s_1,\dots,s_{M}]^{\Trans}\in\{0,1\}^{M}$ where  $s_m=1$ if  molecule mixture $m$ is released otherwise $s_m=0$. We assume that the release times of the Txs are not synchronous or coordinated. We further assume that the molecules propagate independently from each other and the signaling molecules do not react with each other \cite{jamali2019channel}. As channel output,  we consider the aggregated signal of the receptors of the same type, e.g., the output signals of the glomeruli. Let $\by=[y_1,\dots,y_R]^{\Trans}\in\sRp^R$ denote the received signal vector, where $y_r$ is the aggregated output signal of the type $r$ receptors. The exact physical meaning of $y_r,\,\,\forall r$, depends on the adopted Rx and may be the rate of the generated spikes (as in natural olfaction) \cite{kaupp2010olfactory} or the electrical signal generated by a transducer (as in artificial olfaction) \cite{barbosa2018protein}.   The Rx takes multiple  samples with sample interval $\Delta t$, where we use $j$ to denote the sample index. We focus on a short time window of the received signal during which each mixture may be released only once. This leads to the following communication model to relate channel input and output:
\begin{IEEEeqnarray}{lll}
	\text{Release:}\quad &\bu(t) = \sum_{m=1}^M N_{\rls} \delta_{\rm d}(t-\tau_m)\bM_{\rm tx}\bs_m\label{Eq:Release}\vspace{-2.5mm}\\ 
\text{Propagation:}\quad &\bx[j] = {\rm Pois}\big(\bar{\bx}[j]\big)
\quad\text{with}\quad\bar{\bx}[j]=\int_{t=(j-1)\Delta t}^{j\Delta t} \int_{\ell=0}^t \bV(\ell)\bu(t-\ell) {\rm d}\ell\mathrm{d}t \label{Eq:Propagation} \\	
\text{Reception:}\quad	&\by[j]\,\,  = f\big(\bA \bx[j] +\bn[j]\big). \label{Eq:Reception}
\end{IEEEeqnarray}
The variables in \eqref{Eq:Release},  \eqref{Eq:Propagation}, and \eqref{Eq:Reception} corresponding to the release, propagation, and reception mechanisms are respectively explained in the following.

\textbf{Asynchronous release model:} In \eqref{Eq:Release}, $\bu(t) \in\sRp^{Q}$ denotes the vector of released molecule rates at time $t$ and $\delta_{\rm d}(\cdot)$ denotes the Dirac delta function and models the instantaneous release of molecules. Moreover, $\tau_m, m\in\sM$, is the release time of molecule mixture $m$ which is assumed to be random to model asynchronous transmission. To accommodate different random release times, transmit vector $\bs$ is decomposed as $\bs=\sum_{m=1}^M \bs_m$, where $\bs_m\in\{0,1\}^{M}$ is an all-zero vector except for the $m$-th entry which is one if mixture $m$ is released within the considered observation window. Moreover, $\bM_{\rm tx}\in[0,1]^{|\sQ|\times|\sM|}$ is called the molecule-mixture construction matrix, where $[\bM_{\rm tx}]_{q,m}$ determines the fraction of molecule mixture $m$ that is composed of type $q$ molecules. Therefore,  $N_{\rls}$ is the total number of  molecules that each released molecule mixture contains. 
We note that sparsity property \textbf{P2} can be enforced by assuming each molecule mixture contains only few types of molecules and the releases of molecule mixtures by the Txs are  sporadic, i.e., the $\tau_m$ for different $m$ are well separated.

\textbf{Propagation model:} We use  $\bx
[j] \in\sRp^{Q}$ to denote the vector of the \textit{total number} of molecules reaching the sensing volume of the Rx (e.g., the mucus, cf. Fig.~\ref{Fig:Olfactory})  within sampling interval $j$.  We assume that the molecules within the sensing volume either activate a receptor or are degraded so that they cannot activate receptors in the subsequent sampling intervals\footnote{For instance, there might be intermediate biotransformation processes, with signaling molecules being metabolized prior to interaction with the receptors within the mucus layer such that the \textit{sensed} molecules are degraded or even eliminated due to metabolization. This leads to the removal or termination of the sensory stimulation at the receptor site \cite{buettner2017springer}.}. This leads to a model that is conceptually similar to the absorbing Rx model widely adopted in the literature, for which the number of received molecules is known to follow a Poisson distribution when $N_{\rls}$ is large \cite{jamali2018diffusive}. Moreover, in \eqref{Eq:Propagation}, $\bV(t) = \diag(v_1(t),\dots,v_Q(t))\in\sRp^{Q}$ is a diagonal matrix with diagonal entry $v_q(t)$ denoting the channel response for type $q$ molecules, namely  $v_q(t){\rm d} t$ with ${\rm d}t\to 0$ is the probability that a molecule of type $q$ reaches the Rx sensing area in time interval $[t, t+{\rm d}t]$  after being released at $t=0$. The specific shape of $v_q(t)$ depends on the propagation mechanism (e.g., diffusion and laminar or turbulent flow), the Tx-Rx distance, and the properties of the signaling molecules (e.g., their size) \cite{jamali2019channel}.


\textbf{Olfactory-inspired reception model:} In \eqref{Eq:Reception},  $\by[j]$ denotes the  received signal vector at the $j$-th sampling time, where its $r$-th entry is the aggregated  signal of all type $r$ receptors/sensors. Moreover, $\bA\in\sR^{R\times Q}$ denotes the receptor-molecule affinity matrix, where $[\bA]_{r,q}$ determines the strength of the aggregated signal generated by all type $r$ receptors in the presence of a unit concentration of  type $q$ molecules. The cross-reactive property \textbf{P1} is manifested in multiple non-zero elements in each row and column of $\bA$. Examples  of measurement results for the affinity matrix can be found in \cite[Fig.~6]{malnic1999combinatorial} and \cite[Fig.~4]{sicard1984receptor} for a given set of odor molecules and ORs and  in \cite[Fig.~5]{janzen2006colorimetric} for a given set of VOCs  and artificial optical sensors.  Moreover, $f(\cdot)$ is the receptor activation function that is applied to each element of the input vector and is used to model the concentration-dependent non-linearity of the array, cf. property \textbf{P3}. For example, if $f(\cdot)$ is chosen as a rectified linear unit (ReLU) activation function or a SmoothReLU (softplus) activation function, it can model property \textbf{P3} by generating a signal only if the concentration is above a certain threshold. In this paper, we focus on a simple ReLU activation function, namely
\begin{IEEEeqnarray}{lll}  \label{Eq:SoftPlus}
	f(x) = \begin{cases}
		x-x_{\thr} ,\quad &\text{if}\,\,x\geq x_{\thr} \\
		0, &\text{if}\,\,x< x_{\thr},
		\end{cases}
\end{IEEEeqnarray}
where $x_{\thr}$ denotes the receptor activation threshold. The property of inhibition \textbf{P4} can be incorporated in \eqref{Eq:Reception} by assuming that some entries of $\bA$ assume negative values, see \cite[Fig.~6]{pfister2020odorant} for an example characterization of inhibitory/excitatory odor-OR affinity.  A sample construction method for $\bA$ is provided in Section~\ref{sec:A}.  Moreover, $\bn[j]=[n_1[j],\dots,n_R[j]]^\Trans$ $\in\sRp^R$ denotes the noise vector modeling property \textbf{P5},  where $n_r[j]$ is the aggregated random baseline noise of all type $r$ receptors at sample time $j$, which is assumed to be independent of the molecule concentrations $\bx[j]$ and follows a Poisson distribution with mean $\lambda_r$, i.e., $n_r[j]\sim{\rm Pois}(\lambda_r)$ \cite[Remark~21]{jamali2019channel}.  Finally, we note that  the sensitivity and SNR enhancement can be controlled by the relative values of the elements of $\bA$ and $\bn[j]$, i.e., property \textbf{P6}.

\begin{remk}
Models similar to \eqref{Eq:Reception}	have been proposed in the literature \cite{pearce2003chemical,qin2019optimal}. For instance, the authors of \cite{pearce2003chemical} employ a linear model for sensor array optimization, \cite{qin2019optimal} uses nonlinear function $f(x)=\sfrac{x}{(1+x)}$ to model the limited dynamic range of sensors, and \cite{bolding2017complementary} considers binary OR activation which can be model by choosing $f(\cdot)$ as a step function. Nonetheless, as these models were not  developed for communication system design, effects such as multiple and asynchronous releases of molecules by different Txs and inter-symbol and multi-user interference were not accounted for.    
\end{remk}


\subsubsection{\textbf{Sample affinity matrix construction}}\label{sec:A} A key parameter determining the performance of a cross-reactive array is the affinity matrix $\bA$, which plays a similar role as the measurement matrices  widely used in the compressive sensing literature \cite{arjoune2018performance,duarte2011structured}. Instead of focusing on a particular affinity matrix (such as those described in \cite[Fig.~6]{malnic1999combinatorial}, \cite[Fig.~4]{sicard1984receptor}, and \cite[Fig.~5]{janzen2006colorimetric}), we follow the  approach that is common in the compressive sensing literature and \textit{construct} an affinity matrix that complies with the discussed olfaction properties, please refer to   \cite{arjoune2018performance,duarte2011structured} for an overview of structured/unstructured random/deterministic constructions of measurement matrices. The constructed affinity matrix may be used for guiding the choice of molecule and receptor types. The construction methods discussed in \cite{arjoune2018performance,duarte2011structured} were developed for conventional communication systems and do not meet the specific constraints of  affinity matrices of  olfactory systems. Therefore, in the following, we present an algorithm for the construction of a semi-random measurement matrix that adheres to the specific properties of olfaction discussed in Section~\ref{sec:olfprop}. The proposed design follows the following considerations:
\begin{itemize}
	\item For simplicity, we assume that the maximum entry in each column of $\bA$ is normalized to one, i.e., $\max_r a_{q,r}=1,\,\,\forall q$. In other words, for each type of molecule, the type of receptor that generates the strongest signal has normalized affinity one.   
	\item To account for property \textbf{P1}, we assume that each type of molecule activates $R_{\rm act}$ randomly-chosen receptor types. 
	\item To allow for property \textbf{P4}, we assume that the entries of $\bA$ are randomly drawn from interval $a_{q,r}\in[-a_{\rm inh},1],\,\,\forall q,r$, where  negative values imply molecule inhibition. Here, parameter $a_{\rm inh}\in[0,1]$ controls the strength of the maximum inhibition.
	\item The $q$-th column of $\bA$, denoted by $\ba_{q}\in\sR^{R}$, determines how information regarding the presence of molecule type $q$ is distributed across the receptor array. Therefore, the mutual coherence between any two rows of $\bA$, denoted by $\frac{|\ba_q^\Trans\ba_{q'}|}{\|\ba_q\|\|\ba_{q'}\|}$, should be as small as possible in order to avoid the adoption of molecule types that generate similar activation patterns at the receptor array.
\end{itemize} 
Based on the above considerations, the construction of the affinity matrix $\bA$ is summarized in Algorithm~\ref{Alg:Measurement}, where $\mu_{\rm ch}$ is the maximum coherence allowed between the columns of $\bA$. An example of a constructed affinity matrix with $R=10$ receptor types, $M=20$ molecule types, and parameters $a_{\rm inh}=0.3$, $\mu_{\rm thr}=0.5$, and $R_{\rm act} = 5$ is given as follows:
\begin{IEEEeqnarray}{rll}\label{Eq:A}\fontsize{6}{4}\selectfont
	\bA\!=\!
	\arraycolsep=1.3pt\def\arraystretch{2}
	\begin{array}{c}
		\text{R1}\\\text{R2}\\\text{R3}\\\text{R4}\\\text{R5}\\\text{R6}\\\text{R7}\\\text{R8}\\\text{R9}\\\text{R10}
	\end{array}	  	
	\overset{\arraycolsep=5.3pt\def\arraystretch{2}
		\begin{array}{cccccccccccccccccccc}
			~\text{Q1} & ~\text{Q2} & ~\text{Q3} & ~\text{Q4} & ~\text{Q5} & ~\text{Q6} & ~\text{Q7} & ~\text{Q8} & ~\text{Q9} & \text{Q10} & \text{Q11} & \text{Q12} & \text{Q13} & \text{Q14} & \text{Q15} & \text{Q16} & \text{Q17} & \text{Q18} & \text{Q19} & \text{Q20} 
		\end{array}
	}{ 
		\left[\arraycolsep=1.3pt\def\arraystretch{2}
		\begin{array}{cccccccccccccccccccc}
			0 & 0 & 0 & 0 & 0 & 0.55 & 1 & -0.1 & 0 & 0 & 0 & -0.28 & 0.46 & 0.66 & 1 & 0 & 0.76 & -0.14 & 0 & 0 \\ 
			0 & -0.06 & 0.31 & 0.02 & 1 & 0 & 0 & 0.38 & 0.38 & -0.29 & 0 & 0 & 1 & 0 & 0 & 0 & 0.81 & 0.99 & 0 & 0 \\ 
			0 & 0 & 1 & 0.52 & 0.38 & 0.6 & 0 & -0.11 & 0 & 1 & 0 & 0 & 0 & 0 & 0 & 0.01 & 0.98 & 0 & 0 & 0 \\ 
			1 & 0.41 & 0 & 0 & 0 & 0 & 0.27 & 0 & 0.9 & 0 & -0.25 & 0.65 & 0 & -0.25 & 0 & 0 & 1 & 0 & 1 & 0.76 \\ 
			0 & 0 & 0 & 1 & 0 & 0 & -0.01 & 0 & -0.25 & 0 & 0.71 & -0.17 & 0.73 & 0 & 0.38 & 0 & 0 & -0.1 & 0.88 & 0.79 \\ 
			0.55 & 0.44 & 0.55 & 0 & -0.25 & 0.29 & 0 & 1 & 0 & 0 & 0.31 & 0 & 0 & -0.24 & 0.96 & 0.63 & -0.24 & 0 & 0 & 0 \\ 
			-0.3 & 1 & 0 & 0 & 0.5 & -0.29 & 0 & 0 & 0.33 & 0.6 & 0 & 0 & 0 & 1 & 0.12 & -0.17 & 0 & 0 & -0.07 & 0.75 \\ 
			0 & 0 & 0.62 & 0 & 0 & 0 & 0 & -0.19 & 1 & -0.17 & 1 & 0 & -0.2 & -0.13 & 0.4 & 0.55 & 0 & 0 & 0 & 0.36 \\ 
			-0.08 & 0.67 & 0 & 0 & 0 & 1 & -0.21 & 0 & 0 & 0 & 0.45 & 1 & 0 & 0 & 0 & 0 & 0 & 1 & 0.77 & 0 \\ 
			0.16 & 0 & -0.3 & 0.16 & 0.83 & 0 & 0.89 & 0 & 0 & 0.16 & 0 & 0.84 & -0.18 & 0 & 0 & 1 & 0 & 0 & -0.21 & 1
		\end{array} 
		\right] 
	}	\quad\quad\,\,
\end{IEEEeqnarray}

\begin{algorithm}[t] 
	\caption{\small Construction of the Affinity Matrix $\bA$}\scriptsize
	\textbf{input:} $\#$ receptor types $R$, $\#$ molecule types $Q$, inhibition threshold $a_{\rm inh}$, and mutual coherence threshold $\mu_{\rm thr}$ \newline 
	\textbf{output:} Affinity matrix $\bA$
	\begin{algorithmic}[1]\label{Alg:Measurement}
		\FOR{$q=1,\dots,Q$}		
		\STATE Generate $\hat{\ba} = [\hat{a}_1,\dots,\hat{a}_R]^\Trans$ with $R_{\rm act}$ randomly selected entries $\hat{a}_r$ being uniformly distributed RVs in $(0,1]$ and the remaining being zero.
		\STATE Compute $\bar{\ba}= [\bar{a}_1,\dots,\bar{a}_R]^\Trans$ with $\bar{a}_r = \frac{\hat{a}_r}{\max_{r'} \hat{a}_{r'}}(1+a_{\rm inh})-a_{\rm inh}$ if $\hat{a}_r\neq 0$ and $\bar{a}_r =0$ otherwise.
		\STATE Compute $\mu={\max}_{q'<q} \frac{|\bar{\ba}^\Trans\ba_{q'}|}{\|\bar{\ba}\|\|\ba_{q'}\|}$.
		\IF{$q\neq 1$ OR $\mu>\mu_{\rm thr}$}		
		\STATE Go to line~2.
		\ELSE
		\STATE Set $\ba_q=\bar{\ba}$.
		\ENDIF	
		\ENDFOR 		 
	\end{algorithmic}
\end{algorithm}

\section{Transmitter and Receiver Designs}\label{sec:TxRx}

\subsection{Modulation Design} \label{sec:Tx}

In this section, we specify how the Txs encode information into the releases of mixtures of different types of molecules. To this end, we first formally define our notion of  molecule mixture, and then introduce the proposed MMSK modulation design.


\subsubsection{\textbf{MMSK modulation alphabet}} Let $\sQ_k$  denote the set of different types of molecules employed by Tx~$k$,  and $\sM_k$ denote the set of  molecule mixtures constructed from $\sQ_k$ for MMSK modulation.  Our objective is to systematically design the MMSK modulation alphabet (i.e., which mixtures are used at each Tx for signaling).  Ideally,  one may adopt all $2^Q$ possible mixtures for signaling; however, these molecule mixtures are not equally distinguishable by the Rx. In fact, depending on the array architecture, e.g., matrix $\bA$, some mixtures may generate similar activation patterns across the receptor array which are then difficult to discriminate. In a synthetic MC system, we have the privilege to be able to choose which mixtures are most beneficial  for communication. Therefore, mixtures that cannot be well discriminated by the Rx should be excluded from the set of mixtures $\sM_k$ employed for MMSK modulation. To do so, we require a metric which quantifies how dissimilar different mixtures of molecules are.  Unlike in vision and audition, where stimuli can be systematically characterized by a wave frequency, there is no equivalent metric for olfaction and simple features such as the number of carbon atoms and functional groups do not fully describe the response patterns of different chemicals. In \cite{haddad2008metric}, the authors developed a multidimensional odor space model where each odorant is described by 1664 molecular descriptor values. However, this characterization is independent of the odor discrimination capacity of the Rx. In contrast, in this work, we employ a metric that characterizes the difference in structure and composition of the molecule mixtures from the perspective of the Rx array. In particular, let $d_{\bq_1,\bq_2}\big(\bar{\bx}_1,\bar{\bx}_2\big)$ be a metric that quantifies the dissimilarity of two molecule mixtures, which contain the molecule types and  expected concentrations specified by $(\bq_1,\bq_2)$ and $(\bar{\bx}_1,\bar{\bx}_2)$, respectively. In the following, we present the proposed MMSK modulation design for a general metric $d_{\bq_1,\bq_2}\big(\bar{\bx}_1,\bar{\bx}_2\big)$.  

\begin{defin}[$d_{\thr}$-Distinguishable MMSK Modulation Alphabet]\label{def:MMSK} 
	 A set of mixture molecules $\sM$ is called $d_{\thr}$-distinguishable w.r.t. metric $d_{\bq_1,\bq_2}\big(\bar{\bx}_1,\bar{\bx}_2\big)$ if 
	\begin{IEEEeqnarray}{lll}  \label{Eq:MMSKconstraint}
		d_{\bq_{m},\bq_{m'}}\big(\bar{\bx}_{m},\bar{\bx}_{m'}\big) \geq d_{\thr},\quad\forall m\neq m'\in \sM,
	\end{IEEEeqnarray}
\end{defin}
where $\bar{\bx}_{m},\,\forall \sM$, are the \textit{expected} concentrations of the molecules at the Rx.  \hfill $\square$

The constraint in \eqref{Eq:MMSKconstraint} ensures that the adopted mixtures are distinguishable at the Rx.

%

\begin{figure}[t]
	
	\begin{minipage}{0.46\textwidth}
	\centering
\includegraphics[width=1\columnwidth, angle =0]{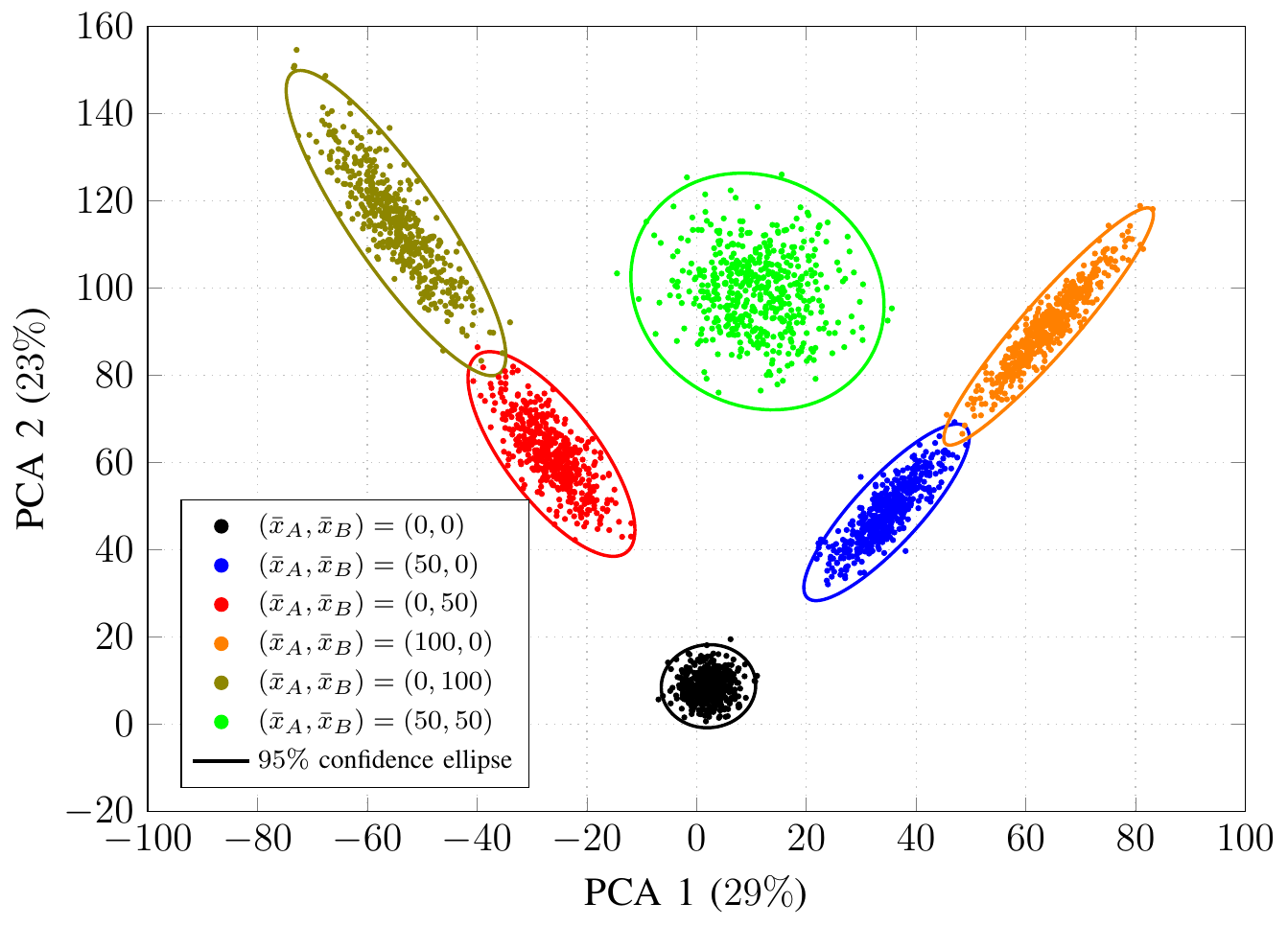}
\vspace{-0.6cm}
\caption{The first two principal components  which account for $29\%$ and $23\%$ of the total data variance.  The results are generated for two types of molecules, denoted by type $A$ and type $B$, which correspond to the first two columns  of affinity matrix $\bA$ in \eqref{Eq:A}. Moreover, we assume $\lambda_r=10$, $x_{\rm thr}=5$. The PCA feature extraction matrix is trained for $10^5$ independent realizations of the considered six cases for $(\bar{x}_A,\bar{x}_B)$  \cite{jurs2000computational}. For each scenario, the smallest ellipse containing $95\%$ of the data is also depicted for better data visualization. \vspace{-0.3cm}} \label{Fig:PCA} \vspace{-0.3cm}
	\end{minipage}
	\begin{minipage}{0.01\textwidth}
		\quad
	\end{minipage}
	\begin{minipage}{0.53\textwidth} \vspace{-0.3cm}
	\centering
\includegraphics[width=0.88\columnwidth, angle =0]{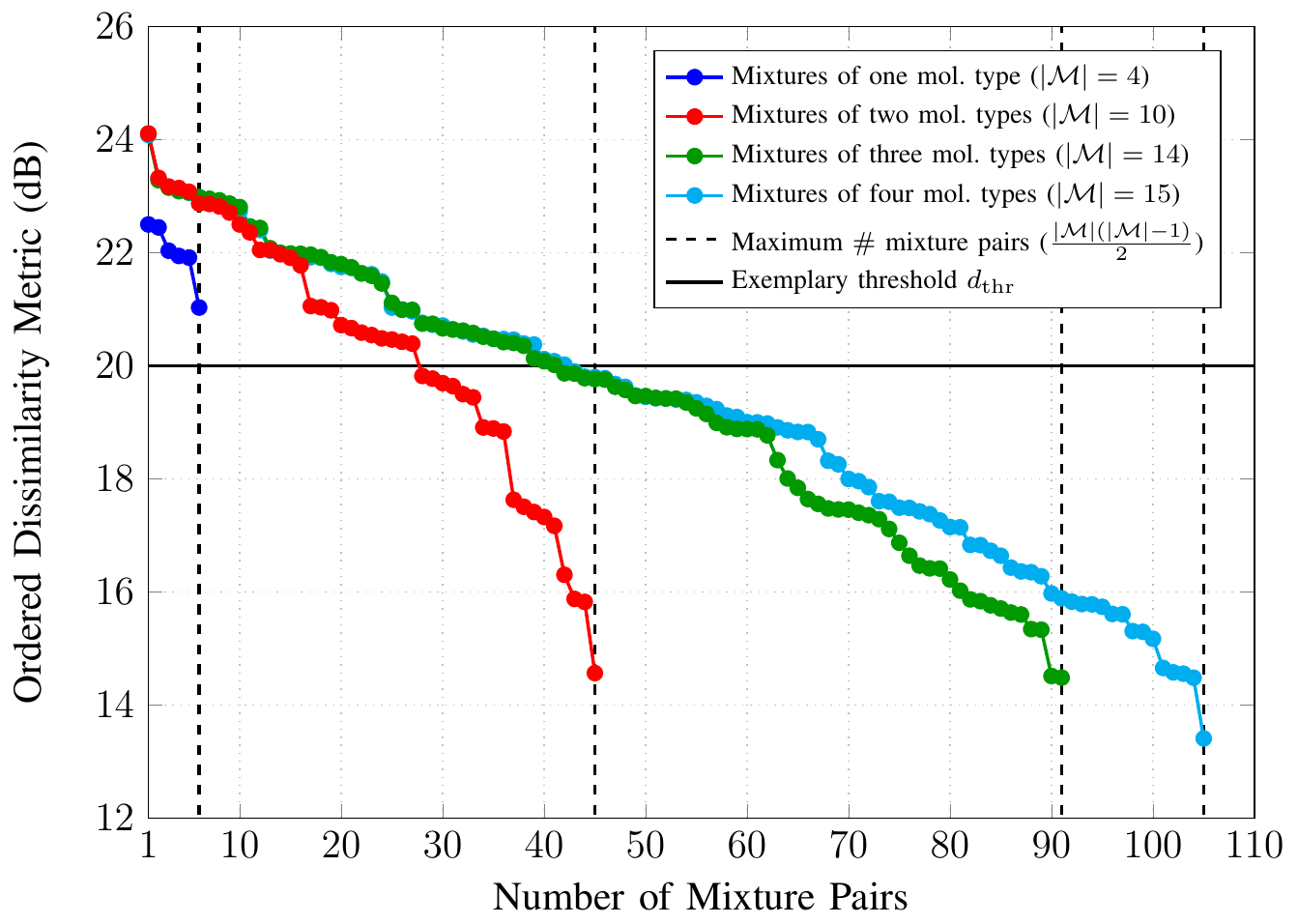}
\vspace{-0.11cm}
\caption{Dissimilarity metric $d_{\bq_{m},\bq_{m'}}\big(\bar{\bx}_{m},\bar{\bx}_{m'}\big)$ (in dB) ordered in a descending order for all possible pairs of mixtures $(\bq_{m},\bq_{m'})$ from set $m,m'\in\sM$. The results are generated for $Q=4$ types of molecules, which are obtained from  four randomly-chosen columns (columns $1, 4, 8, 15$) of affinity matrix $\bA$ in \eqref{Eq:A}. Moreover, we assume $\lambda_r=10$, $x_{\rm thr}=5$, and uniform molecule distribution for all mixtures such that $\sum_{q}\bar{x}_q=100$ holds. The results are averaged for $10^4$ realizations of RVs $\bx_m$ and $\bn$. \vspace{-0.3cm}} \label{Fig:SNR_mix} \vspace{-0.3cm}
	\end{minipage}
	
\end{figure} 

\subsubsection{\textbf{Proposed dissimilarity metric}} Before formally presenting the proposed metric, we first establish the underlying rationale for a simple scenario. Let us adopt  affinity matrix $\bA$ in \eqref{Eq:A} (i.e., $R=10$ types of receptors and $Q=20$ types of molecules) and choose the molecules corresponding to the first two columns of $\bA$ as type $A$ and type $B$ molecules whose expected concentrations at Rx are denoted by $\bar{x}_A$ and $\bar{x}_B$, respectively. We further assume $\lambda_r=10$ and $x_{\rm thr}=5$. In order to visualize the received signal vector of the Rx array, $\by$, we apply principal  component analysis (PCA) to $\by$ as a dimensionality reduction technique \cite{jurs2000computational}. In Fig.~\ref{Fig:PCA}, we plot the first two principal components of the extracted PCA features for several concentrations of type $A$ and type $B$ molecules. First of all, this figure shows that different concentrations of the same type of molecule (e.g., $(\bar{x}_A,\bar{x}_B)=(50,0)$ and $(100,0)$) can be difficult to distinguish whereas the molecule mixture $(\bar{x}_A,\bar{x}_B)=(50,50)$ is more easily distinguishable from the other considered cases. This illustrates the potential benefits of  MMSK modulation compared to CSK and MSK modulation. In addition, Fig.~\ref{Fig:PCA} suggests that the randomness of the received signal is not the same in different directions. Therefore, the dissimilarity metric should not only include the relative distance between two mixture vectors in the receptor space but also the relative variation along the line connecting these two clusters which is relevant for potential mis-classification of the two mixtures by the Rx.  Motivated by this observation, we define the proposed dissimilarity metric as follows
\begin{IEEEeqnarray}{lll}  \label{Eq:metric}
	d_{\bq_{m},\bq_{m'}}\big(\bar{\bx}_{m},\bar{\bx}_{m'}\big)  = \frac{\left\|\Ex\{\by_{m}\}-\Ex\{\by_{m'}\}\right\|_2^2}{\Vx\{\bp_{m,m'}^\Trans \by_{m}\}+\Vx\{\bp_{m,m'}^\Trans \by_{m'}\}},
\end{IEEEeqnarray}
where $\bp_{m,m'}=\frac{\Ex\{\by_{m}\}-\Ex\{\by_{m'}\}}{\left\|\Ex\{\by_{m}\}-\Ex\{\by_{m'}\}\right\|_2}$ and the expectations are w.r.t. the random molecule concentration $(\bx_m,\bx_{m'})$ and noise $\bn$. In other words, the metric proposed in \eqref{Eq:metric} is a generalized SNR quantity defined in a high-dimensional space of the Rx array. Thereby,  the numerator specifies the expected Euclidean distance between the received signals of molecule mixtures $m$ and $m'$, whereas the denominator determines the variance of the signal fluctuations along direction $\bp_{m,m'}$.

Fig.~\ref{Fig:SNR_mix} illustrates the metric $d_{\bq_{m},\bq_{m'}}\big(\bar{\bx}_{m},\bar{\bx}_{m'}\big)$ proposed in \eqref{Eq:metric} for an example scenario where different mixtures are generated based on five types of molecules. In the following, we highlight several interesting design intuitions that can be drawn from this figure. First, we observe from Fig.~\ref{Fig:SNR_mix} that not all pairs of molecule mixtures are equally separable. In fact, for mixtures composed of a larger number of molecule types, there are more mixture pairs that are difficult to distinguish at the Rx (e.g., pairs for which $d_{\bq_{m},\bq_{m'}}\big(\bar{\bx}_{m},\bar{\bx}_{m'}\big)\leq d_{\rm thr}$ holds). Therefore, we have to avoid the simultaneous presence of such mixtures in the modulation alphabet of one Tx. Interestingly, Fig.~\ref{Fig:SNR_mix} also suggests that mixtures of molecule types may be more separable than non-mixtures (i.e., only one molecule type). For the example considered in Fig.~\ref{Fig:SNR_mix}, the most separable pairs of mixtures are $(\bq_{m},\bq_{m'})=(4,8)$ with $d_{\bq_{m},\bq_{m'}}\big(\bar{\bx}_{m},\bar{\bx}_{m'}\big)=22.45$~dB for one type of molecule, and $(\bq_{m},\bq_{m'})=([1,8],[4,15])$ with $d_{\bq_{m},\bq_{m'}}\big(\bar{\bx}_{m},\bar{\bx}_{m'}\big)=24.15$~dB when the mixtures may contain any of the four considered types of molecules. This is due to the fact that the reception space of the Rx array is not orthogonal w.r.t. different types of molecules. Hence, single types of molecules do not necessarily correspond to the maximum separation and larger separation can be achieved by choosing proper molecule mixtures. At the same time, the achievable separation saturates for  more than two molecule types in the mixtures.

\subsubsection{\textbf{Molecule-Tx allocation and molecule-mixture alphabet construction}} Assume that there are $Q$ different types of molecules available for communication between $K$ Txs and the Rx. Motivated by the discussion in the previous subsections, next, we first present an algorithm  which allocates these types of molecules to the $K$ Txs, i.e., determining $\sQ_k,\,\,\forall k$. Subsequently, given set $\sQ_k$, we develop another algorithm to construct for  Tx $k$  the corresponding MMSK modulation alphabet $\sM_k,\,\,\forall k$.

\begin{algorithm}[t] 
	\caption{\small Molecule-Tx Allocation}\scriptsize
	\textbf{input:} Dissimilarity metric $d_{m,m'}\big(\bx_{m},\bx_{m'}\big)$ for all pairs of molecule types $m,m'\in\sQ$; $\#$ of Txs $K$, and $\#$ molecule types per Tx $Q_{\rm tx}$. \newline 
	\textbf{output:} Molecule set allocated to each Tx, i.e., $\sQ_k,\,\,\forall k$.	
	\begin{algorithmic}[1]\label{Alg:Mol_Tx}
		\STATE Set $\sQ_k=\emptyset,\,\,\forall k$, and $\widehat{\sQ}=\sQ$. \hfill \textit{$\%$ Initialization}
		\FOR{$k=1,\dots,K$}	 
		\STATE Find $(q^*,q'^{*})=\underset{q,q'\in\widehat{\sQ}}{\argmax}\,\, d_{q,q'}\big(x_{q},x_{q'}\big)$. \hfill \textit{$\%$ Molecule pair selection}
		\STATE Update $\sQ_k\leftarrow\sQ_k\cup\{q^*,q'^{*}\}$ and $\widehat{\sQ}\leftarrow\widehat{\sQ}\backslash\{q^*,q'^{*}\}$. \hfill \textit{$\%$ Initial molecule-pair allocation}
		\ENDFOR 
		\WHILE{$\exists k: \,\,|\sQ_k|<Q_{\rm tx}$}
		\STATE Find $k^*=\underset{k:\,|\sQ_k|<Q_{\rm tx}}{\argmax} \,\, \underset{q\in\widehat{\sQ}}{\max}\,\,\underset{q'\in\sQ_k}{\min}\,\, d_{q,q'}\big(x_{q},x_{q'}\big)$. \hfill \textit{$\%$ Tx selection}
		\STATE Find $q^*= \underset{q\in\widehat{\sQ}}{\argmax}\,\,\underset{q'\in\sQ_{k^*}}{\min}\,\, d_{q,q'}\big(x_{q},x_{q'}\big)$. \hfill \textit{$\%$ Molecule selection}
		\STATE Update $\sQ_{k^*}\leftarrow\sQ_{k^*}\cup\{q^*\}$ and $\widehat{\sQ}\leftarrow\widehat{\sQ}\backslash\{q^*\}$. \hfill \textit{$\%$ Sequential molecule allocation}
		\ENDWHILE		 
	\end{algorithmic}
\end{algorithm}

\textbf{Molecule-Tx allocation:} As discussed in  Section~\ref{sec:channel}, the Txs are assumed to sporadically access the MC channel. Hence, the multi-user interference is expected to be small in the considered scenario. This motivates us to allocate molecules such that the molecules assigned to a given Tx have the largest possible separation because they will be used to construct one MMSK alphabet. In addition, we assume a non-overlapping allocation of molecule types to Txs (i.e., ${\cap}_k\sQ_k=\emptyset$, where $\emptyset$ denotes the empty set). This allows the identification of the transmitting Tx at the Rx based on received signal vector $\by$ without requiring an extra identification protocol. Note that finding the optimal allocation is a combinatorial problem in $M$, i.e., there  are $\prod_{k=1}^K {{M-(k-1)M_{\rm tx}}\choose{M_{\rm tx}}}$ possible cases with $M_{\rm tx}$ denoting the number of molecules allocated per Tx. Therefore, a full search is computationally challenging for $M\gg1$. Instead, we propose a greedy algorithm, summarized in Algorithm~\ref{Alg:Mol_Tx}, which proceeds sequentially and in each step assigns the molecule that  maximizes  the dissimilarity metric w.r.t. the set of previously allocated molecules (i.e., lines 6-10 of Algorithm~\ref{Alg:Mol_Tx}). Moreover, for initialization, we allocate the $K$ non-overlapping pairs of molecules with maximum pair-wise dissimilarity metric to the $K$ Txs, respectively (i.e., lines 2-5 of Algorithm~\ref{Alg:Mol_Tx}). 

As an example, applying Algorithm~\ref{Alg:Mol_Tx} to  affinity matrix $\bA$ given  in \eqref{Eq:A} (i.e., $Q=20$ molecule types), assuming $K=4$, $Q_{\rm tx}=4$, and the remaining parameters given in the caption of Fig.~\ref{Fig:SNR_mix}, we obtain the worst- and best-case values for the dissimilarity metric (in dB) as $(22.2,23.1)$, $(22.5,22.9)$, $(22.5,22.8)$, and $(22.4,22.8)$ among the molecule types allocated to Txs 1, 2, 3, and 4,  respectively.

\begin{algorithm}[t] 
	\caption{\small Construction of $d_{\rm thr}$-distinguishable MMSK Modulation Alphabet}\scriptsize
	\textbf{input:} Dissimilarity metric $d_{m,m'}\big(\bx_{m},\bx_{m'}\big)$ for all pairs of molecule mixtures $m,m'\in\sM_k^{\rm full}$ and threshold $d_{\rm thr}$. \newline 
	\textbf{output:} $d_{\rm thr}$-distinguishable MMSK modulation alphabet $\sM_k$.	
	\begin{algorithmic}[1]\label{Alg:MMSK}
		\STATE Find $(m^*,m'^{*})=\underset{m,m'\in\sM_k^{\rm full}}{\argmax}\,\, d_{\bq_m,\bq_{m'}}\big(\bx_{m},\bx_{m'}\big)$. \hfill \textit{$\%$ Initial mixture pair selection}
		\IF{$d_{\bq_{m^*},\bq_{m'^{*}}}\big(\bx_{m^*},\bx_{m'^*}\big)<d_{\rm thr}$}
		\STATE Set $\sM_k=\emptyset$  and terminate the algorithm. \hfill \textit{$\%$ A $d_{\rm thr}$-distinguishable does not exist}
		\ELSE 
		\STATE Set $\sM_k = \{m^*,m'^{*}\}$. \hfill \textit{$\%$ Initialization of the MMSK alphabet}
		\ENDIF
		
		\REPEAT 
		\STATE Find $(m^*,m'^*)= \underset{m\in\sM_k^{\rm full}\backslash\sM_k}{\argmax}\,\,\underset{m'\in\sM_{k}}{\min}\,\, d_{\bq_m,\bq_{m'}}\big(\bx_{m},\bx_{m'}\big)$. \hfill \textit{$\%$ Mixture selection}
		\IF{$d_{\bq_{m^*},\bq_{m'^{*}}}\big(\bx_{m^*},\bx_{m'^*}\big)<d_{\rm thr}$}
		\STATE Terminate the algorithm. 
		\ELSE 
		\STATE Update $\sM_k \leftarrow \sM_k \cup\{m^*\}$. \hfill \textit{$\%$ Sequential MMSK alphabet expansion}
		\ENDIF
     	\UNTIL{$\sM_k^{\rm full}\backslash\sM_k=\emptyset$}		 
	\end{algorithmic}
\end{algorithm}

\textbf{MMSK alphabet construction algorithm:} Given $\sQ_k,\,\,\forall k$, obtained from Algorithm~\ref{Alg:Mol_Tx}, next, we design a $d_{\rm thr}$-distinguishable MMSK modulation alphabet $\sM_k$, cf. Definition~\ref{def:MMSK}. Let $\sM_k^{\rm full}$ denote the set of all possible mixtures at Tx $k$ that contain a maximum of $M_{\rm mix}$ molecule types. The maximum number of mixture constituents $M_{\rm mix}$ may be chosen smaller than $|\sQ_k|$ due to complexity considerations. Thereby, in general, there are $2^{|\sM_k^{\rm full}|}-1=\prod_{m=1}^{M_{\rm mix}} 2^{{|\sQ_k|}\choose{m}}-1$ possible non-empty MMSK modulation alphabets, which makes a full search for finding $d_{\rm thr}$-distinguishable MMSK modulation alphabets computationally demanding even for small numbers of molecules $|\sQ_k|$.  To cope with this issue, we propose a greedy algorithm in Algorithm~\ref{Alg:MMSK}, which successively enlarges the MMSK alphabet until adding an additional mixture would violate the $d_{\rm thr}$-distinguishability condition (i.e., lines 7-14 of Algorithm~\ref{Alg:MMSK}). Moreover, for initialization, we choose that pair of mixtures that has the maximum  dissimilarity metric (i.e., lines 1-6 of Algorithm~\ref{Alg:MMSK}).

Using the same example as used before for Algorithm~\ref{Alg:Mol_Tx} and applying Algorithm~\ref{Alg:MMSK} to $\sQ_1=\{1,5,11,14\}$ obtained from Algorithm~\ref{Alg:Mol_Tx} with $M_{\rm mix}=3$ and $d_{\rm thr}=20$~dB, Algorithm~\ref{Alg:MMSK} constructs out of $\sum_{m=1}^{M_{\rm mix}} {{|\sQ_1|}\choose{m}}=14$ possible mixtures the MMSK alphabet $\sM_1=\{\{1,5\},\{1,11\},\{5,14\},\{5,11\},$ $\{11,14\}\}$ with a minimum dissimilarity metric among mixture pairs of $20.68$ dB$\geq d_{\rm thr}$. For this specific example, all mixtures of one and three molecules and mixture $\{1,14\}$ of two molecules are excluded from the MMSK alphabet, cf. Table~\ref{table:MMSK} and Section~\ref{sec:sim} for a more detailed discussion.

\subsection{Recovery Problem}\label{sec:Rx}
Given the modulation design developed in the previous section, we now explain how the Rx recovers the sent messages (or equivalently the corresponding molecule mixtures) based on the observation from cross-reactive array. In the following, we first develop a recovery problem for one received signal sample, i.e., $\by[j]$. Moreover, we present the concept of adaptive recovery where the parameters of the proposed recovery problem are adaptively chosen. Finally, we introduce matched filtering to combine the outputs of the proposed recovery problem for multiple sampling times.

\subsubsection{\textbf{Single-sample concentration recovery}} Motivated by the sparsity of $\bs$ (or equivalently $\bx$), we formulate the recovery task as a compressive sensing problem. For rigorousness of presentation, we use $\byrv\in\sRp^{R}$ to denote an RV representing the received signal where $\by$ denotes one realization of  $\byrv$. A typical compressive sensing problem formulation is to choose the sparsest $\bx$ that reconstructs a signal close to the observation $\by$ based on the adopted measurement model, e.g., \eqref{Eq:Reception}, namely \cite{blumensath2013compressed,defraene2013declipping}
\begin{IEEEeqnarray}{rll}\label{Eq:CS}
\text{OP0}:	\underset{\bx\in\sR_{\geq0}^{Q}}{\min}\,\, &\|\bx\|_0\nonumber\\
	\text{s.t.}\quad & \big\|\by-\Ex\{\byrv\}\big\|_2^2\leq \epsilon',
\end{IEEEeqnarray}
where $\epsilon'$ is the threshold parameter that controls the reconstruction error and the expectation is w.r.t. noise $\bn$ for given molecule concentration $\bx$. The value of $\epsilon'$ depends on the variance of $\byrv$ for a given $\bx$, which is determined by $\lambda_r$. For simplicity,  throughout this paper, we employ $\epsilon'=\epsilon\Vx\{\byrv\}$, where $\epsilon$ is the normalized reconstruction error threshold.


Problem OP0 faces few shortcomings in general and in particular for the problem under consideration in this paper, which are summarized in the following. \textit{i)} The zero-norm optimization is a non-deterministic polynomial-time (NP) hard problem. This issue is often addressed in the literature by replacing $\|\cdot\|_0$ with its convex one-norm relaxation $\|\cdot\|_1$ \cite{blumensath2013compressed}. \textit{ii)} The constraint in \eqref{Eq:CS} is non-convex due to non-linear receptor activation function $f(\cdot)$. A similar challenge exists for the recovery of clipped audio signals, where a convex reformulation of the clipping function is proposed in \cite{defraene2013declipping} by separating the clipped and un-clipped signals. \textit{iii)}  The problem in  \eqref{Eq:CS} treats all possible $\bx$ similarly and does not exploit the fact that depending on the  MMSK modulation alphabet adopted by the Txs, only certain mixtures may be sent by the Txs.  To address issues \textit{i)} and \textit{ii)}, we adapt the technique used in \cite{defraene2013declipping} to the problem in \eqref{Eq:CS} accounting for the fact that, unlike the Gaussian noise assumed in  \cite{defraene2013declipping}, the interfering noise molecules in our setup are not zero-mean RVs. This leads to the following recovery problem

\begin{equation}   \label{Eq:CSprop_x}
	\begin{aligned}[c]
	\text{OP1:}\underset{\bx\in\sRp^{Q}}{\min}\,\,\, &\|\bx\|_1\\
	\text{s.t.}  \,\,
	\text{C1:}\,\, &\big\|\by_{\cA}-\Ex\{\byrv_{\cA}^{\rm lin}\}\big\|_2^2\leq \epsilon\Vx\big\{\bone_{|\cA|}^\Trans\byrv_{\cA}^{\rm lin}\big\}\\
	        \text{C2:}\,\,  & \Ex\{\byrv_{\cAc}^{\rm lin}\} \leq  x_{\thr}\bone_{|\cAc|}+\sqrt{\epsilon\Vx\big\{\byrv_{\cAc}^{\rm lin}\big\}}
	    \end{aligned}
    \!\!\!\Longrightarrow\!\!\!
    	\begin{aligned}[c]
    	\underset{\bx\in\sRp^{Q}}{\min}\,\,\, &\|\bx\|_1\\
    	\text{s.t.}  \,\,
    	\text{C1:}\,\, &\|\by_{\cA}-(\bA_{\cA}\bx+(\lambda_r-x_{\thr})\bone_{|\cA|})\|_2^2\leq|\cA|\lambda_r\epsilon\\
    	\text{C2:}\,\,  &\bA_{\cAc}\bx+(\lambda_r-x_{\thr})\bone_{|\cAc|} \leq \sqrt{\lambda_r\epsilon}\bone_{|\cAc|}
    \end{aligned}
\end{equation}
where  $\cA$ and $\cAc$ are the sets of activated and non-activated receptors, respectively, $\by_{\cA}\in\sRp^{|\cA|}$ collects the measurements from the activated  receptors,  $\byrv^{\rm lin}\triangleq\bA\bx+\bn$ is an RV that is the input to  activation function $f(\cdot)$ in \eqref{Eq:Reception},  $\byrv^{\rm lin}_{\cA}\in\sRp^{|\cA|}$ and $\byrv^{\rm lin}_{\cAc}\in\sRp^{|\cAc|}$ collect the elements of $\byrv^{\rm lin}$ from the activated and non-activated receptors, respectively, and $\bA_{\cA}\in\sR^{|\cA|\times Q}$ and $\bA_{\cAc}\in\sR^{|\cAc|\times Q}$ are matrices containing only the rows of $\bA$ corresponding to activated and non-activated receptors, respectively. Thereby, constraint C1 limits the reconstruction error over the set of activated receptors. Since the noise is random and unknown, the exact condition $\byrv^{\rm lin}_{\cAc}\leq x_{\thr}\bone_{\cAc}$ cannot be applied for the non-activated receptors. Therefore, we consider the statistical constraint in C2 where the normalized parameter $\epsilon$ controls the strictness of this constraint.


Problem OP1 employs the space of molecule concentrations as the signal recovery space.  In order to cope with issue \textit{iii)}, we choose the space of the concentrations of the molecule mixtures as the signal recovery space. In other words, instead of deciding which types of molecules are present around Rx, we directly determine which mixtures are present around Rx. A similar concept is used to describe  mixture identification by natural olfactory systems \cite{thomas2014perception}, which is known as elemental and configural processing of mixtures, where in the former case, a mixture is identified by its constituents whereas in the latter case, it is identified as a unique quantity.    
To formalize this, let us decompose the \textit{expected} received molecule concentration as $\bar{\bx}=\bM_{\rm rx}\bw$, where $\bw\in\sRp^{M}$ is the vector of expected concentrations of molecule mixtures used by the Txs for signaling. Moreover,  we refer to $\bM_{\rm rx}\in[0,1]^{Q\times M}$ as the molecule-mixture reception matrix, where $[\bM_{\rm rx}]_{q,m}$ determines the \textit{expected} fraction of molecule mixture $m$ observed at the Rx that are type $q$ molecules.  If the propagation properties of all constituent molecules in the mixture are identical, we obtain $\bM_{\rm rx}=\bM_{\rm tx}$. Otherwise, the exact value of $\bM_{\rm rx}$ varies over time and depends on the sampling time and the channel response function $v_q(t),\,\,\forall q$. Since the release times of the mixtures are random and not known at the Rx, we employ the expected fraction of received type $q$ molecules over time, defined as $\gamma_q\triangleq\int_{t=0}^{\infty}v_q(t)\mathrm{d}t$, to construct the following approximation for $\bM_{\rm rx}$
\begin{IEEEeqnarray}{lll}  \label{Eq:Mrx}
	[\bM_{\rm rx}]_{q,m} \approx \frac{[\bM_{\rm tx}]_{q,m} \gamma_q}{\sum_{q=1}^Q [\bM_{\rm tx}]_{q,m} \gamma_q}.
\end{IEEEeqnarray}
Recall that for a given $\bar{\bx}$, the actual number of received molecules is an RV, cf. \eqref{Eq:Propagation}. Therefore, for rigorousness of presentation, we use $\bxrv\in\sRp^{Q}$ to denote an RV representing the vector of received numbers of different molecule types where  $\bx$ denotes one realization of  $\bxrv$.  Using these notations, we propose the following recovery~problem: 
\begin{IEEEeqnarray}{rll}  \label{Eq:CSprop_mix}
		\begin{aligned}[c]
		\text{OP2}:		\underset{\bx\in\sRp^{Q},\bw\in\sRp^{M}}{\min}\quad &\|\bw\|_1\\
		\text{s.t.}  \,\,
		\text{C1}, \text{C2}, 
		\text{C3:}\,\,& |\bx-\Ex\{\bxrv\}|^2\leq \delta\Vx\{\bxrv\}
	\end{aligned}
	\!\!\!\Longrightarrow\,
	\begin{aligned}[c]
	\underset{\bx\in\sRp^{Q},\bw\in\sRp^{M}}{\min}\quad &\|\bw\|_1\\
		\text{s.t.}  \,\,
		\text{C1},\text{C2},
		\text{C3:}\,\,  &|\bx-\bM_{\rm rx}\bw|^2\leq\delta\bM_{\rm rx}\bw,
	\end{aligned}
\end{IEEEeqnarray}
where constraint C3 controls the maximum deviation of the estimated molecule concentration $\bx$ from  the mean of $\bxrv$. The amount of allowable deviation, in general, depends on the variance of $\bxrv$, which, for the Poisson model in \eqref{Eq:Propagation}, is equal to its mean, i.e., $\Ex\{\bxrv\}=\Vx\{\bxrv\}=\bar{\bx}=\bM_{\rm rx}\bw$. Therefore, we parameterized the maximum deviation as $\delta\Vx\{\bxrv\}$, where $\delta$ is a constant threshold to control the deviation. The optimal values of $\delta$ and $\epsilon$ are numerically determined for different setups in Section~\ref{sec:sim}. The optimization problem in \eqref{Eq:CSprop_mix} is convex and can be solved by standard optimization toolboxes, e.g., CVX. For future reference, let $\hat{\bw}$ and $\hat{\bx}$ denote the solution of OP2.

We assume that each Tx releases only one mixture at a time and that, due to the random access by the Txs, the probability that two Txs release mixtures simultaneously is negligible. Therefore, we employ the following peak detector to determine which molecule mixture is present at each sampling~time:
\begin{IEEEeqnarray}{lll}  \label{Eq:ThrDec}
	\hat{s}_{m} = \begin{cases}
	1,	& \mathrm{if}\,\, m=\underset{m'\in\sM}{\argmax}\,\hat{w}_{m'}\\
	0, & \mathrm{otherwise}.
	\end{cases}
\end{IEEEeqnarray}

\subsubsection{\textbf{Adaptive recovery}} Problem OP2 is solved assuming that \textit{any of the molecule mixtures} used for signaling at all Txs may arrive at the Rx at \textit{any time} due to the spontaneous access of Txs to the MC channel.   Therefore, the possible presence of molecule mixtures used by all TXs for signaling is considered in OP2 via matrix $\bM_{\rm rx}$. However, recall that the  sets of molecules allocated to different Txs are non-overlapping, i.e., $\bigcap_{k=1}^K\sQ_k=\emptyset$. Therefore, upon detection of specific types of molecules, the Rx is able to infer which Tx sends information. Since the Txs access the channel sporadically, the knowledge of which Tx is active can be subsequently exploited to improve the recovery performance. In fact, natural olfactory systems are believed to be adaptive and to selectively process odor stimuli \cite{rolls2008selective,kendrick1992changes}.  For example, in \cite{rolls2008selective}, human subjects were instructed to rate the pleasantness and intensity of an odor under two scenarios: \textit{i)} knowing that the odor source is a jasmine flower and \textit{ii)} not knowing the odor source. It was observed that the neural activations in the medial orbitofrontal and pregenual cingulate
cortex (responsible for emotion and memory processing) were greater under scenario \textit{i)} than under scenario \textit{ii)}, which suggests  selective processing of odor stimuli by the human brain. Such adaptive processing can be also incorporated into recovery problem OP2 by adaptively choosing mixture matrix $\bM_{\rm rx}$. Let $\bM_k$ denote a submatrix of $\bM_{\rm rx}$ consisting  only of the columns $m\in\cM_k$ of $\bM_{\rm rx}$, which correspond to the molecule mixtures used by Tx $k$. Using this notation and assuming that Tx $k$ has been already inferred to be active by the Rx, we propose to solve OP2 using $\bM_k$ instead of $\bM_{\rm rx}$ to achieve improved message recovery. We will numerically evaluate the efficiency of the proposed adaptive processing in Section~\ref{sec:sim}. 

\subsubsection{\textbf{Multiple-sample matched filtering}} Due to the strong dispersion of the MC channel, the molecules released by the Txs will reach the Rx over an extended period of time. Therefore, instead of making the decision based on a single sample as in \eqref{Eq:ThrDec}, one may first filter the mixture concentration samples $\hat{w}_m[j]$ with a linear filter, i.e.,   
\begin{IEEEeqnarray}{lll}  \label{Eq:MF}
	\hat{w}^{\rm flr}_m[j] = \sum_{\kappa=-\infty}^\infty f_m[\kappa] \hat{w}_m[j-\kappa],
\end{IEEEeqnarray}
where $\hat{w}^{\rm flr}_m[j]\in\sR$ and $f_m[\kappa]\in\sR,\,\,\forall j$, denote the filtered signal and the filter coefficients for the $m$-th mixture, respectively. We determine the filter to match the channel response of the constituent molecules, i.e., $v_q(t)$,  as follows \cite{jamali2017design}
\begin{IEEEeqnarray}{lll}  \label{Eq:MF_coeff}
	f_m[\kappa]= \sum_{q} c_q v_{q}(-\kappa\Delta t) [\bM_{\rm tx}]_{q,m},
\end{IEEEeqnarray}
where the above summation is over the indices of constituent molecules in mixture $m$ and $c_q=\frac{1}{\sum_{\kappa=0}^\infty v_q^2(\kappa\Delta t)}$ is introduced only for normalization purpose.
The filtered signal $\hat{w}^{\rm flr}_m[j]$ has peaks that correspond to the release times of mixture $m$.

\section{Simulation Results}\label{sec:sim}

In this section, we first present the considered simulation setup. Subsequently, we evaluate the performance of the proposed modulation design and message recovery problems.

  \begin{table}[t]
	\caption{Default values of system parameters.\vspace{-0.15cm}}
	\label{table:parameter}
	\centering
	\scalebox{0.6}{
		\begin{tabular}{||c|c|c||}\hline
			Parameter & Definition & Value \\\hline
			$K$ & $\#$ Txs & $4$\\ \hline
			$R$ & $\#$ receptors & $10$ \\ \hline
			$Q$ & $\#$ molecule types &  $20$ \\ \hline
			$Q_{\rm tx}$ & $\#$ molecule types per Tx &  $4$ \\ \hline
			$M_{\rm tx}$ & MMSK alphabet size &  $4$ \\ \hline 
			$M_{\rm mix}$ & Max. $\#$ molecules in mixture &  $3$ \\ \hline 
			$N_{\rls}$ & $\#$ released molecules &  $10^5$ \\ \hline
			$(\alpha_q,\beta_q,\gamma_q)$ & \makecell{channel parameters \\ (normalized peak time)} &  $(0.5,1.7,0.01)$ \\ \hline
			$\bar{x}_{\rm mix}$ & \makecell{expected $\#$ of received \\ mixture molecules} &  $50$ \\ \hline
			$\lambda_r$ & noise mean &  $10$ \\ \hline
			$x_{\rm thr}$ & activation threshold &  $5$ \\ \hline
			$\bA$ & affinity matrix &  Eq. \eqref{Eq:A}\\ \hline
		\end{tabular}
	}\vspace{-0.4cm}
\end{table}   

\subsection{Simulation Setup}

Unless stated otherwise, the values of the system parameters are chosen as given in Table~\ref{table:parameter}. We assume that each Tx spontaneously accesses the channel and sends one mixture corresponding to one message. The transmission, propagation, and reception of molecules follow  \eqref{Eq:Release}, \eqref{Eq:Propagation}, and \eqref{Eq:Reception}, respectively. To be able to analyze the features of the proposed MC system in detail, we first focus on single-sample recovery in Figs.~\ref{Fig:Pe_opt_rnd}-\ref{Fig:Pe_param}. 
In this case, we assume uniform molecule distribution for all adopted mixtures, i.e., we assume $\sum_{q}\bar{x}_q[j]=\bar{x}_{\rm mix},\,\,\forall j$, holds. Multiple-sample recovery  and the proposed matched filtering are investigated in Figs.~\ref{Fig:COM}a-\ref{Fig:COM}f, where the release and propagation processes are explicitly taken into account.   We note that regardless of the specific propagation mechanism, the MC channel can be characterized in a generic manner by  the fraction of molecules that reach the Rx over time, the speed of molecule propagation, and the speed of the decay of the number of  molecules around the Rx   \cite{jamali2019channel}. Therefore, instead of focusing on a particular propagation mechanism, for the results in Figs.~\ref{Fig:COM}a-\ref{Fig:COM}f, we consider the following parametric model that  captures the aforementioned general features of the MC channel
\begin{IEEEeqnarray}{rll}\label{Eq:channel} 
	v_q(t) = \frac{\gamma_q}{\beta_q}\left[1+\frac{\alpha_q}{\beta_q}\right]\left[1-\e^{-\frac{t}{\alpha_q}}\right]\e^{-\frac{t}{\beta_q}},
\end{IEEEeqnarray}
where $\gamma_q$ is the average fraction of molecules reaching the Rx over  time, i.e., $\gamma_q=\int_{0}^\infty v_q(t) {\rm d} t$, and $\alpha_q$ and $\beta_q$ respectively determine the speed of molecule propagation and decay.


\subsection{Performance Evaluation}
In the following, we study the performance of the proposed message modulation and  recovery designs.

\textbf{MMSK modulation construction:} Table~\ref{table:MMSK} summarizes the results obtained for the molecule-Tx allocation and MMSK modulation design obtained with Algorithms~\ref{Alg:Mol_Tx} and \ref{Alg:MMSK}, respectively, for affinity matrix $\bA$ in \eqref{Eq:A}. The designed mixtures are reported according to the order in which they are generated by Algorithm~\ref{Alg:MMSK}. Thereby, we observe that the minimum value of the dissimilarity metric for the generated mixtures for each Tx (i.e., $\underset{m,m'\in\sM_k}{\min}\,\, d_{\bq_m,\bq_{m'}}\big(\bx_{m},\bx_{m'}\big)$) decreases as the alphabet size increases. This implies that the larger the number of different messages that a Tx has, the worse the overall recovery performance will be. It is interesting to observe that for all Txs the best mixtures contain neither only one type of molecule nor all types of molecules. This is because, on the one hand, increasing the number of molecule types per mixture may lead to more separable mixtures as the signals of the pure molecules do not form an orthogonal basis in the reception space, cf. Fig.~\ref{Fig:PCA}. On the other hand, including too many molecules in each mixture implies  a lower relative concentration per constituent molecule type, which is not desirable. For the example scenario considered here, employing two types of molecules in each mixture is  optimal.   

\begin{table}[t]
	\caption{Molecule-Tx allocation and MMSK modulation design using Algorithms~\ref{Alg:Mol_Tx} and \ref{Alg:MMSK}, respectively, assuming $\bar{x}_{\rm mix}=100$. The mixtures are given according to the order in which they are generated by Algorithm~\ref{Alg:MMSK}.  The corresponding values of the minimum dissimilarity matrix among the modulation alphabets are reported, too. \vspace{-0.15cm}}
	\label{table:MMSK}
	\centering
	\scalebox{0.6}{
		\begin{tabular}{||c|cccc|cccc|cccc|cccc||}\hline
			&\multicolumn{4}{|c}{Tx~1}
			&\multicolumn{4}{|c}{Tx~2}
			&\multicolumn{4}{|c}{Tx~3}
			&\multicolumn{4}{|c||}{Tx~4}\\\hline
		\makecell{Allocated molecules\\ 
			(Alg.~\ref{Alg:Mol_Tx})}
		&\multicolumn{4}{|c}{$\{1,     5,    11,    14\}$}
		&\multicolumn{4}{|c}{$\{3,     7,    12,    19\}$}
		&\multicolumn{4}{|c}{$\{2,     6,    13,    16\}$}
		&\multicolumn{4}{|c||}{$\{9,    10,    15,    18\}$}\\\hline
		\makecell{Designed mixtures\\ 
			(Alg.~\ref{Alg:MMSK})}			
		&\multicolumn{3}{|c}{Mixtures}&\makecell{Min. Dis. \\ Met. (dB)}
		&\multicolumn{3}{|c}{Mixtures}&\makecell{Min. Dis. \\ Met. (dB)}
		&\multicolumn{3}{|c}{Mixtures}&\makecell{Min. Dis. \\ Met. (dB)}
		&\multicolumn{3}{|c}{Mixtures}&\makecell{Min. Dis. \\ Met. (dB)}\\\hline
		Mixture~1 & - & 5 & 14 & - & - & 7 & 12 & - & - & 2 & 16 & - & - & 9 & 18 & - \\ 
		Mixture~2 & - & 1 & 11 & 25.14 & - & 3 & 19 & 24.75 & - & 6 & 13 & 24.68 & - & 10 & 15 & 24.91 \\ 
		Mixture~3 & - & 1 & 5 & 21.44 & - & 3 & 7 & 21.19 & - & 2 & 13 & 21.22 & - & 9 & 10 & 21.36 \\ 
		Mixture~4 & - & 11 & 14 & 21.18 & - & 7 & 19 & 20.64 & - & 2 & 6 & 21.02 & - & 10 & 18 & 21.21 \\ 
		Mixture~5 & - & 5 & 11 & 20.72 & - & 3 & 12 & 20.53 & - & 13 & 16 & 20.88 & - & 15 & 18 & 21.18 \\ 
		Mixture~6 & - & 1 & 14 & 19.93 & - & 12 & 19 & 19.87 & - & 6 & 16 & 20.47 & - & 9 & 15 & 21.05 \\ 
		Mixture~7 & - & - & 1 & 17.17 & - & - & 7 & 17.47 & - & - & 2 & 17.44 & - & - & 10 & 17.71 \\ 
		Mixture~8 & - & - & 14 & 16.59 & - & - & 3 & 17.06 & - & - & 16 & 17.28 & 9 & 15 & 18 & 17.08 \\ 
		Mixture~9 & 5 & 11 & 14 & 16.58 & 3 & 7 & 12 & 16.59 & 2 & 13 & 16 & 17.08 & - & - & 18 & 17.05 \\ 
		Mixture~10 & - & - & 5 & 16.21 & 3 & 7 & 19 & 16.21 & - & - & 13 & 16.96 & - & - & 9 & 16.91 \\ 
		Mixture~11 & 1 & 5 & 11 & 16.15 & - & - & 19 & 16.01 & 6 & 13 & 16 & 16.89 & 10 & 15 & 18 & 16.82 \\ 
		Mixture~12 & - & - & 11 & 15.87 & - & - & 12 & 15.9 & 2 & 6 & 13 & 16.05 & - & - & 15 & 16.52 \\ 
		Mixture~13 & 1 & 5 & 14 & 15.34 & 7 & 12 & 19 & 15.22 & - & - & 6 & 15.95 & 9 & 10 & 18 & 16.48 \\ 
		Mixture~14 & 1 & 11 & 14 & 15.27 & 3 & 12 & 19 & 15.03 & 2 & 6 & 16 & 14.88 & 9 & 10 & 15 & 16.47 \\  \hline
		\end{tabular}
	}
\end{table}   

\begin{figure}[t]
	
	\begin{minipage}{0.495\textwidth}
	\centering
\includegraphics[width=1\columnwidth, angle =0]{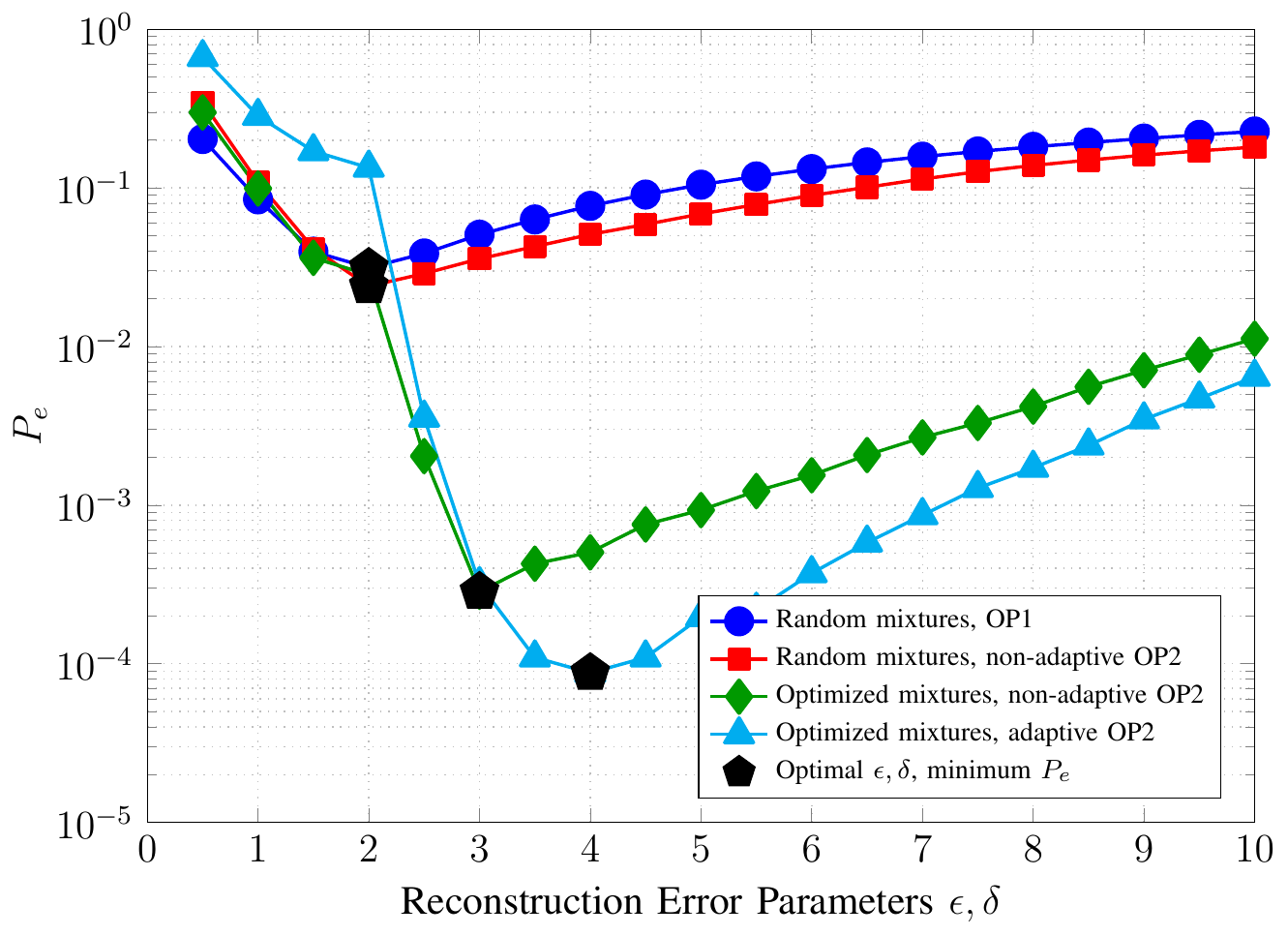}
\vspace{-0.5cm}
\caption{Error probability $P_e$ vs. reconstruction error parameters $\epsilon,\delta$ for $K=4$ Txs and  $M_{\rm tx}=4$ mixtures per Tx of two constituent molecules chosen either randomly or from Table~\ref{table:MMSK} (i.e., optimized).\vspace{-0.3cm}} \label{Fig:Pe_opt_rnd}\vspace{-0.1cm}
	\end{minipage}
	\begin{minipage}{0.01\textwidth}
		\quad
	\end{minipage}
	\begin{minipage}{0.495\textwidth} 
	\centering
\includegraphics[width=1\columnwidth, angle =0]{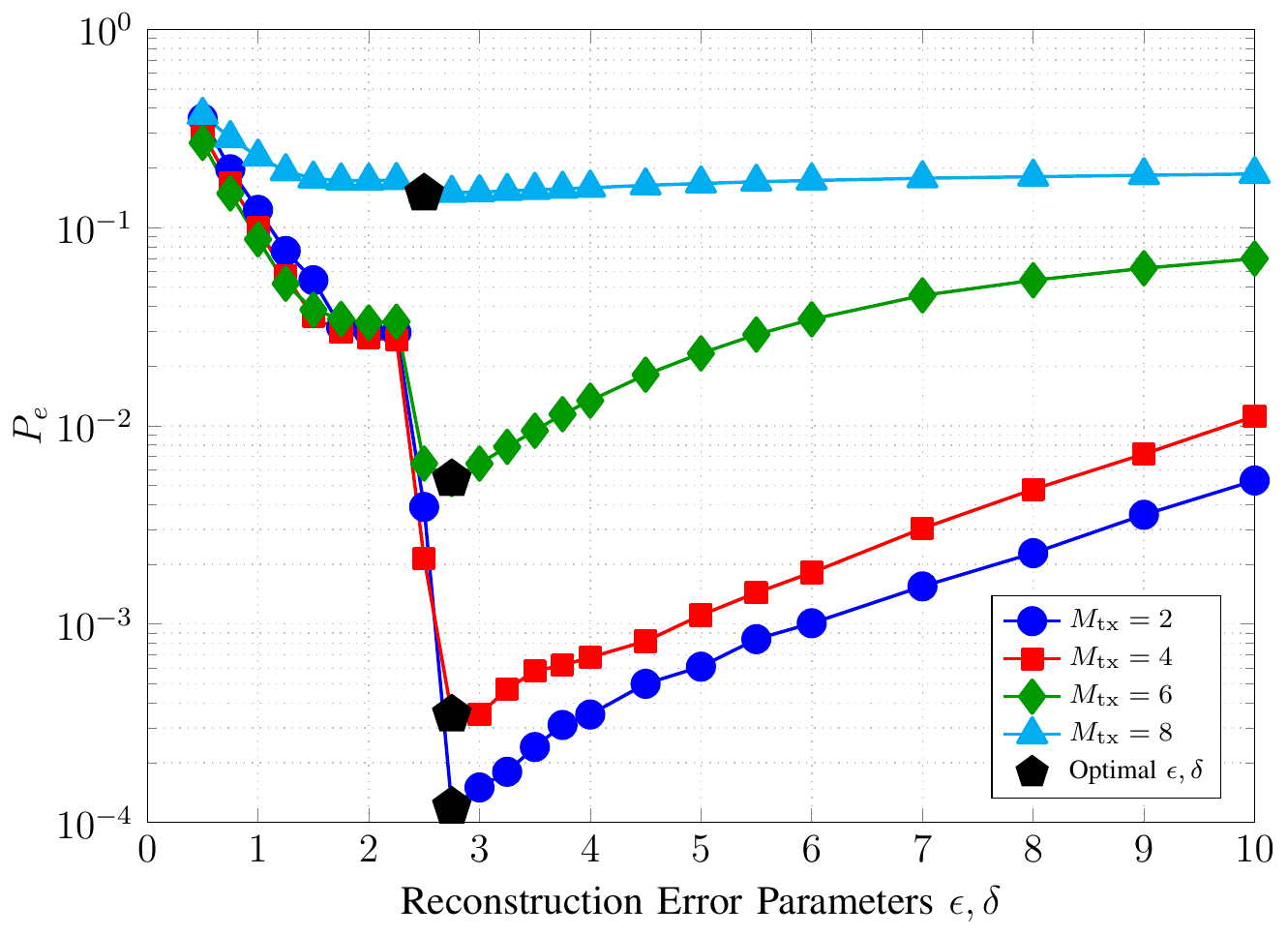}
\vspace{-0.5cm}
\caption{Error probability $P_e$ vs. reconstruction error parameters $\epsilon,\delta$ for $\sM_k$ chosen from Table~\ref{table:MMSK}, different $M_{\rm tx}=|\cM_k|$, and non-adaptive recovery problem OP2.\vspace{-0.3cm}} \label{Fig:Pe_known}\vspace{-0.1cm}
	\end{minipage}
	
\end{figure}

\textbf{Random vs. optimized and adaptive vs. non-adaptive  recovery:} In Fig.~\ref{Fig:Pe_opt_rnd}, we plot the probability of erroneous molecule mixture detection, defined as $P_e = \Pr\big(\hat{\bs} \neq \bs\big)$, vs. the reconstruction error parameters $\epsilon,\delta$ for different mixtures of two molecules chosen either randomly or optimized, cf. Table~\ref{table:MMSK}. Since both reconstruction error parameters are similarly normalized, cf. \eqref{Eq:CSprop_mix}, we assume identical values for them, i.e., $\epsilon=\delta$. For the proposed recovery problem OP2, we show the results for both adaptive and non-adaptive processing depending on whether or not the identity of the active Tx is inferred.  We observe from Fig.~\ref{Fig:Pe_opt_rnd} that for each curve, there exists an optimal value for the reconstruction error parameter that minimizes the error probability. This is due to the fact that while large  $\epsilon,\delta$ lead to inaccurate estimates $\hat{\bx}$ and $\hat{\bw}$ (i.e., underfitting), small  $\epsilon,\delta$ lead to an infeasible recovery problem and overfitting. Moreover, we observe a significant performance improvement (two orders of magnitude) when optimized mixtures are used for the Tx's MMSK alphabets. Furthermore, knowledge of the active Tx considerably improves the performance of the proposed adaptive recovery problem OP2. Finally, Fig.~\ref{Fig:Pe_opt_rnd} suggests that as more knowledge about the received signal (i.e., the MMSK  modulation and/or transmitting Tx) is exploited for mixture recovery, the optimal value of the reconstruction error parameters $\epsilon,\delta$ increases. This is because, in this case, larger $\epsilon,\delta$ can be accommodated without considerable underfitting since the feasible set is already reduced by exploiting the extra knowledge.

\textbf{Impact of MMSK alphabet size on recovery performance:} The size of the MMSK alphabet depends on the number of different messages that the Tx wishes to convey to the Rx. In Fig.~\ref{Fig:Pe_known}, we plot the error probability vs. the reconstruction error parameter for different numbers of mixtures $M_{\rm tx}$ (i.e., MMSK alphabet size). We show results for recovery problem OP2 without exploiting the knowledge of the transmitting Tx.  It can be observed from this figure that as the number of mixtures increases, the error probability increases. This behavior stems from the fact that by increasing $M_{\rm tx}$, the minimum value of the dissimilarity metric among the mixtures in the alphabet decreases, cf. Table~\ref{table:MMSK}, which constitutes the bottleneck for  the recovery performance. 

%

\begin{figure}[t]
	\centering
	\includegraphics[width=0.495\columnwidth, angle =0]{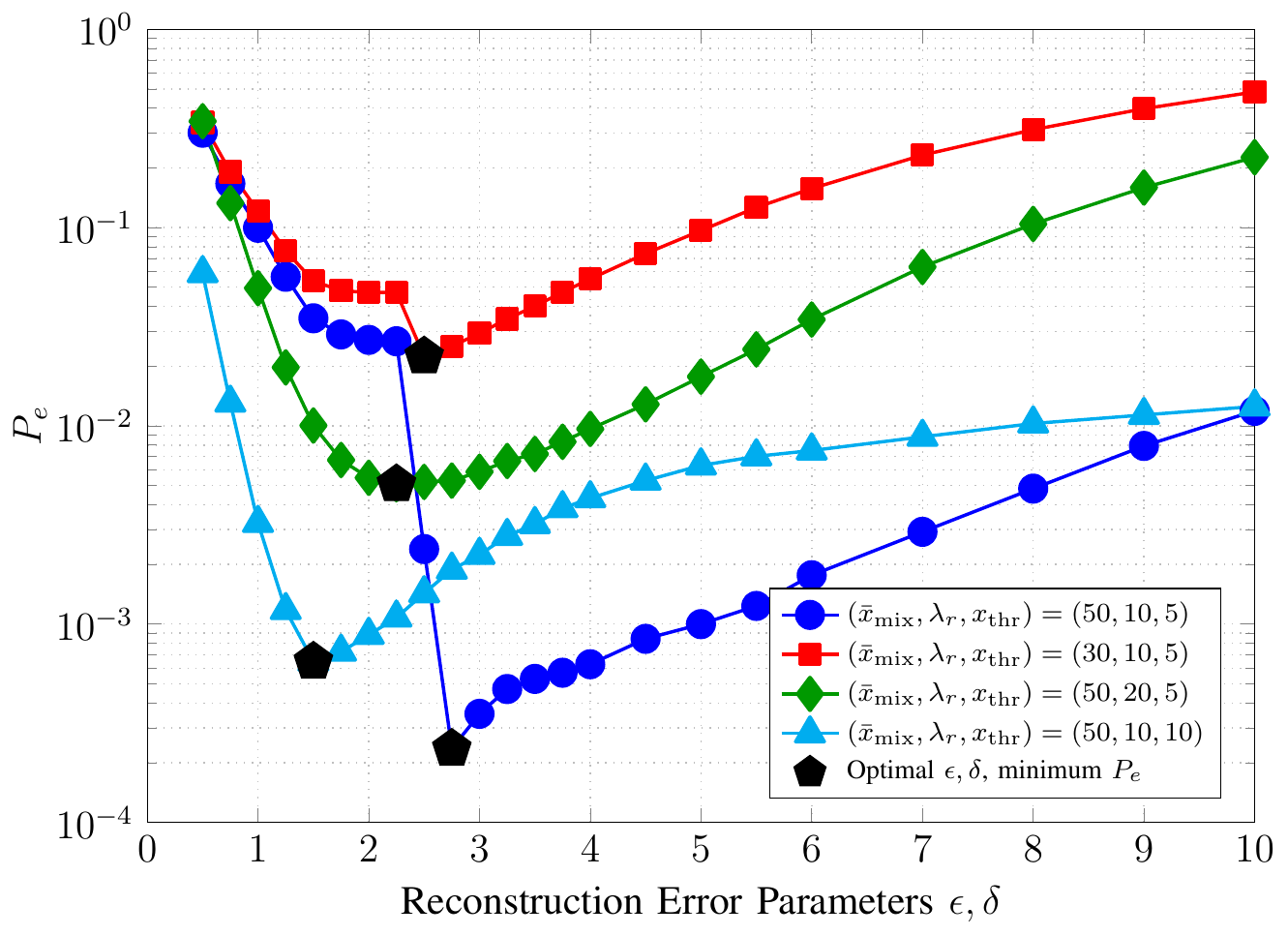}
	\vspace{-0.1cm}
	\caption{Error probability $P_e$ vs. reconstruction error parameters $\epsilon,\delta$ for $\sM_k$ chosen from Table~\ref{table:MMSK}, different values of the system parameters $(\bar{x}_{\rm mix},\lambda_r,x_{\rm thr})$, and non-adaptive recovery problem OP2. \vspace{-0.3cm}} \label{Fig:Pe_param} \vspace{-0.3cm}
\end{figure} 

\textbf{Impact of system parameters on the recovery performance:} In Fig.~\ref{Fig:Pe_param}, we show the error probability vs. the reconstruction error parameters for different values of system parameters $(\bar{x}_{\rm mix},\lambda_r,x_{\rm thr})$. It can be observed from this figure that an increase in the concentration of the received mixtures (i.e., $\bar{x}_{\rm mix}$) significantly improves the recovery performance. There are two reasons for this performance improvement. First, a higher concentration implies a higher SNR and reduces the impact of noise. Second, higher concentrations of molecules  activate even receptors with low affinity, which implies that more knowledge about the present mixtures is contained in the receptor array signal.  This argument is further confirmed by Fig.~\ref{Fig:Pe_param} as the recovery performance improves for smaller  noise means $\lambda_r$ and smaller activation thresholds~$x_{\rm thr}$.

\begin{figure} 
	\centering 
	\begin{minipage}[t]{0.49\columnwidth}
		\includegraphics[width=1\columnwidth, angle =0]{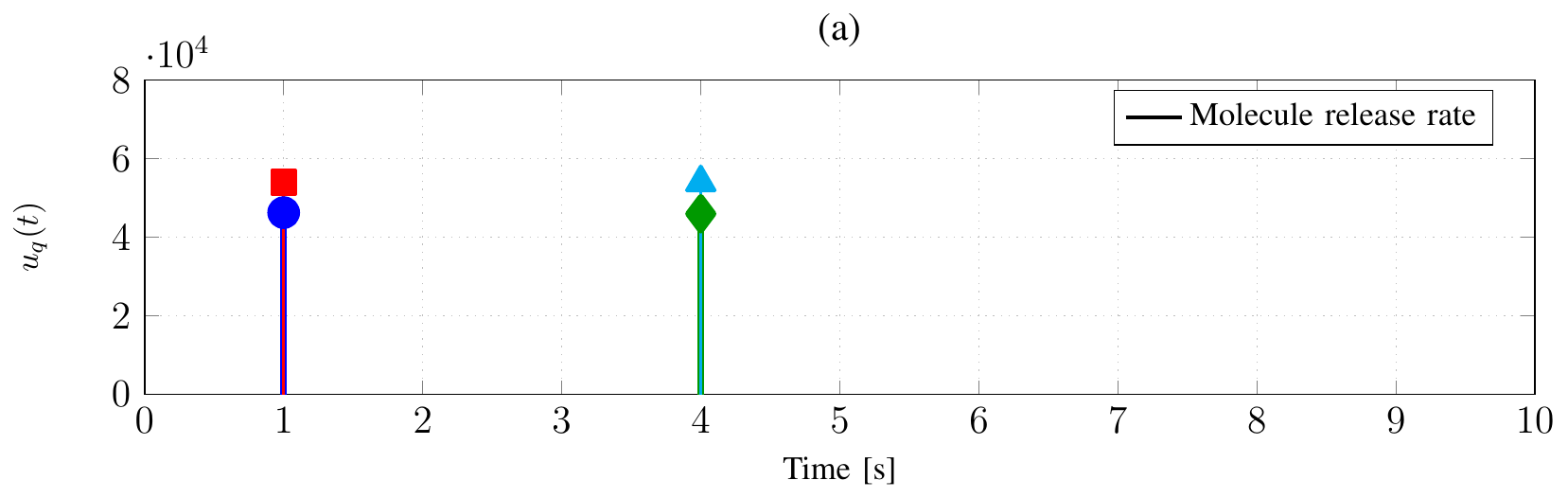}
		\includegraphics[width=1\columnwidth, angle =0]{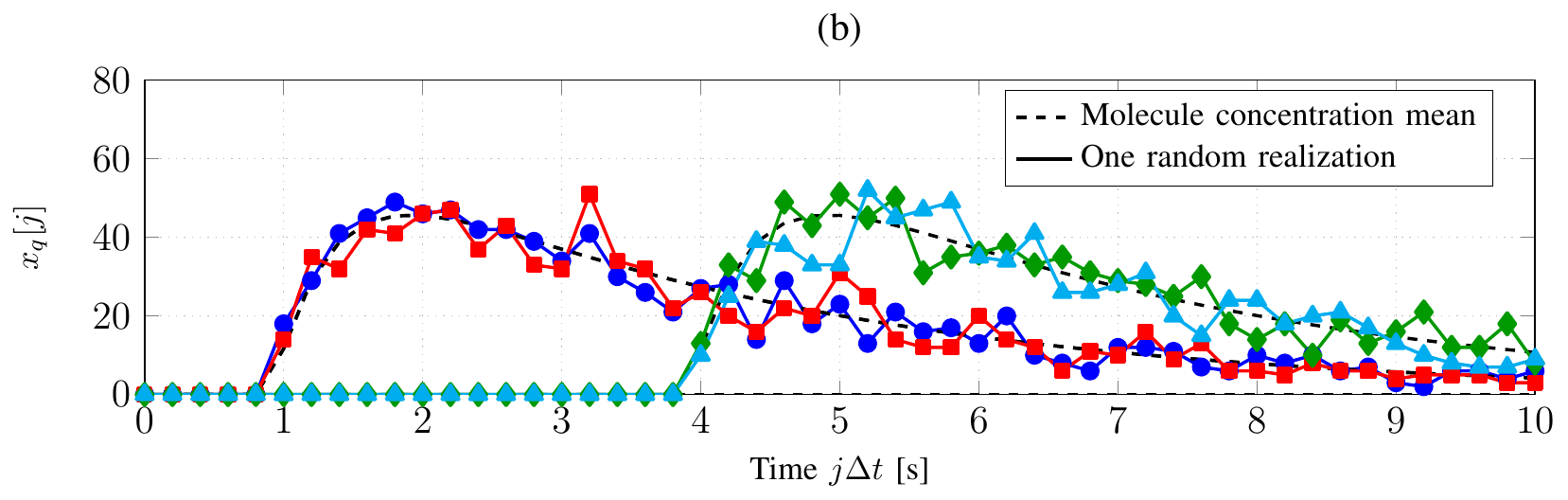}
		\includegraphics[width=1\columnwidth, angle =0]{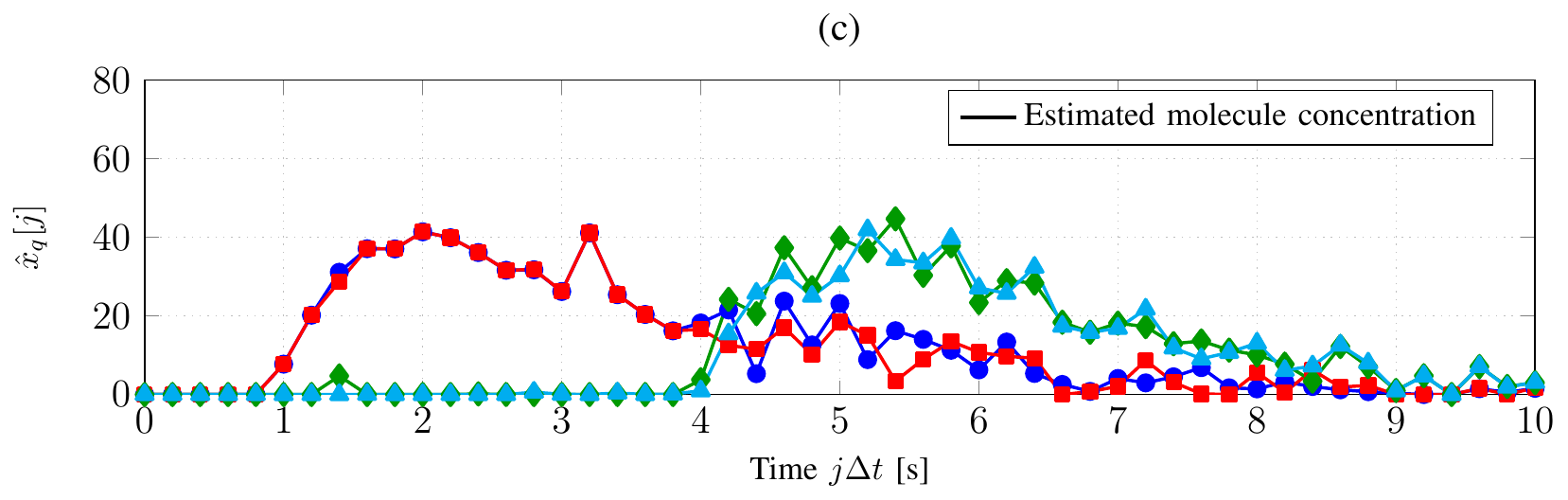}
	\end{minipage}
	\begin{minipage}[t]{0.49\columnwidth}
		\includegraphics[width=1\columnwidth, angle =0]{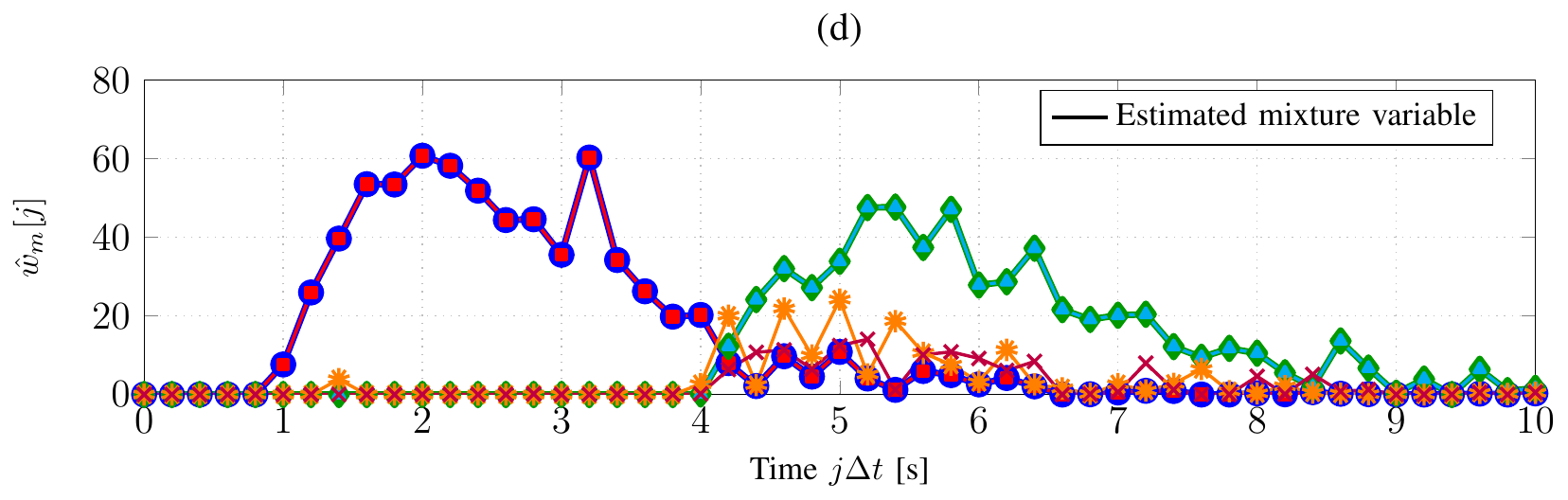}
		\includegraphics[width=1\columnwidth, angle =0]{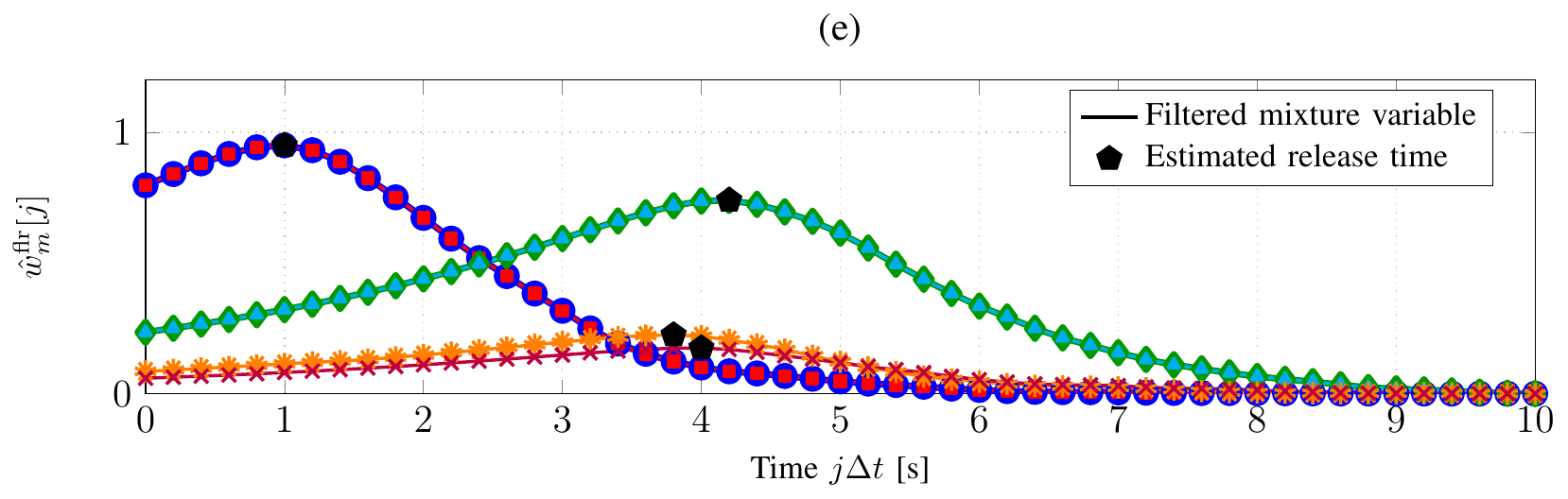}
		\includegraphics[width=1\columnwidth, angle =0]{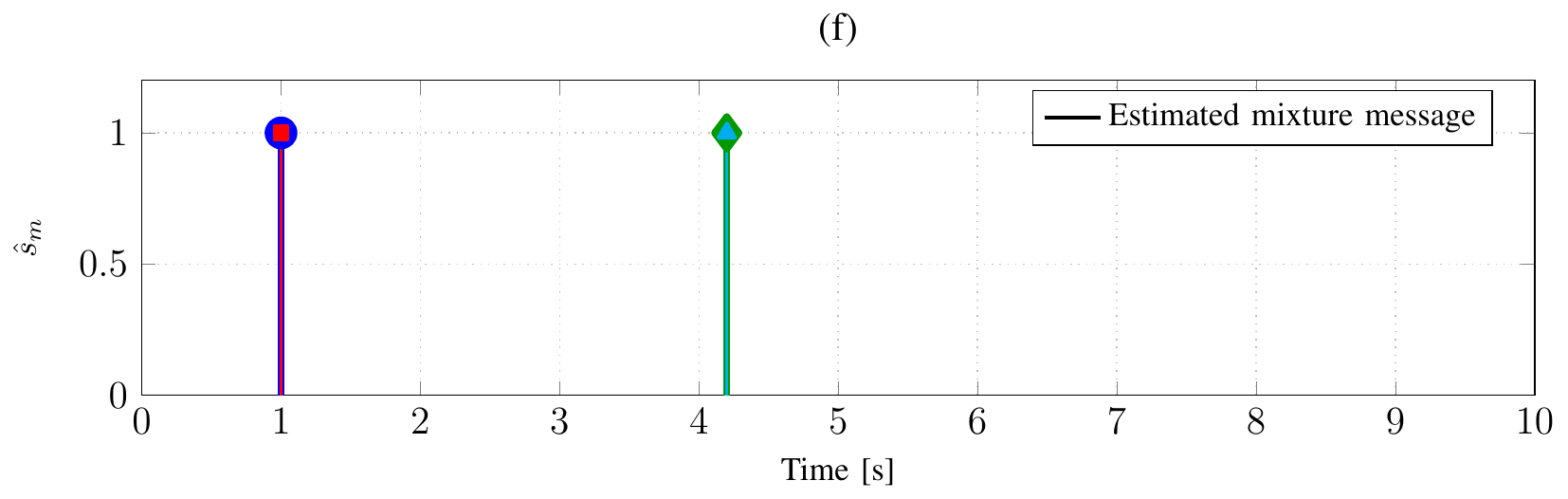}
	\end{minipage}
	\caption{Illustration of the signals occurring in the proposed communication system during  release, propagation, reception, and post-processing. The four released types of molecules are shown as blue circle, red square, green diamond, and cyan triangle, respectively.  \vspace{-0.3cm}} \label{Fig:COM} 
\end{figure}

\textbf{Recovery of multiple samples via matched filtering:} So far, we studied single-sample recovery based on OP2. Next, we investigate the benefits of combining multiple samples using the proposed matched filter, cf. \eqref{Eq:MF}, \eqref{Eq:MF_coeff}. Fig.~\ref{Fig:COM} shows the signals occurring in the considered communication system during molecule release, propagation, reception, and post-processing, as explained in the following.

\begin{itemize}
	\item Fig.~\ref{Fig:COM}(a) presents the molecule release rates $u_q(t)$ corresponding to two mixtures each having two constituent molecules (namely the first two mixtures of Tx~1 reported in Table~\ref{table:MMSK}, where molecules 5, 14, 1, and 11 are shown as blue circle, red square, green diamond, and cyan triangle, respectively). The height of the Dirac delta function schematically shows the number of released molecules, i.e., $N_{\rm rls}/2$. 
	\item Fig.~\ref{Fig:COM}(b) shows the number of molecules that reach the sensing volume of the Rx within each sample time, i.e., $x_q[j]$, for a sampling duration of $\Delta t = 0.2$~s. As can be seen from this figure, $x_q[j]$ is random and extends over a long time period due to the dispersion of the molecular propagation process, where significant interference among two consecutive releases of mixtures is clearly~visible.
	\item Fig.~\ref{Fig:COM}(c) shows the  molecule concentration $\hat{x}_q[j]$ estimated via OP2 with adaptive recovery assuming $M_{\rm tx}=4$, i.e., Tx~1 uses four signaling mixtures. We observe from this figure that $\hat{x}_q[j]$ can follow the temporal dynamics of $x_q[j]$ despite the inherently high degree of randomness. Moreover, since we employ adaptive recovery (i.e., $\bM_1$ is used in OP2 instead of $\bM_{\rm rx}$), the search space is limited to only the molecules that Tx~1  uses (namely molecules 5, 14, 1, and 11). 
	\item Fig.~\ref{Fig:COM}(d) reports the estimated mixture variable $\hat{w}_m[j]$. We show the two transmitted mixtures (namely [5, 14] and [1, 11]) with the colors and markers of their constituent molecules whereas the non-transmitted mixtures are shown with new colors/markers (i.e., orange star for mixture [1, 5] and purple cross for mixture [11, 14]). We see from this figure that while problem OP2 mostly correctly distinguishes the two transmitted mixtures. Nevertheless, at some sampling instances (especially for low concentrations, e.g., at $t\approx 5$~s), the presence of a low concentration of other mixtures may (erroneously) better explain the received observation. 
	\item Fig.~\ref{Fig:COM}(e) shows the filtered signal $\hat{w}^{\rm flr}_m[j]$ in \eqref{Eq:MF}. We see from this figure that while the unfiltered signal $\hat{w}_m[j]$ looks still quite random due to the local randomness of the received signal, the filtered signal $\hat{w}^{\rm flr}_m[j]$ is much smoother as the randomness is averaged over time. Moreover, $\hat{w}^{\rm flr}_m[j]$ has clearly identifiable peaks that correspond to the release times of the mixtures. 
	\item  Fig.~\ref{Fig:COM}(f)  finally shows the recovered mixture data using the peak detector in  \eqref{Eq:ThrDec}. The transmitted mixtures (i.e., messages) are not only correctly identified but also their release times are estimated almost perfectly within the sampling precision of $\Delta t = 0.2$~s.
\end{itemize}

\section{Conclusions}
We have proposed and studied  a novel olfaction-inspired  synthetic MC system. In the proposed MC system, Txs employ mixtures of molecules as information carriers and the Rx is equipped with an array of cross-reactive receptors in order to identify the transmitted mixtures. We first introduced an end-to-end MC channel model that accounts for the key properties of olfaction. Given a set of signaling molecule types, we then developed algorithms that allocate each Tx different molecule types and construct favorable molecule mixtures (i.e., the MMSK modulation alphabet) for communication. Subsequently, we formulated the molecule mixture recovery as a convex compressive sensing problem, i.e., OP2, which explicitly takes into account knowledge of the MMSK modulation alphabets employed by the Txs. Our simulation results reveal the high performance gains achieved with the proposed optimized MMSK modulation compared to non-optimized mixtures and the high detection capability of the proposed mixture recovery based on problem OP2 for various sets of system parameters. 

\bibliographystyle{IEEEtran}
\bibliography{References}

\end{document}